\documentclass[a4paper,11pt]{article}
\usepackage[utf8]{inputenc}
\usepackage{lmodern}
\usepackage[T1]{fontenc}
\usepackage{amsmath,amsthm,amsfonts,amssymb,bm}
\usepackage{dsfont}     
\usepackage{enumitem}
\usepackage{graphicx}
\usepackage{bbm}
\usepackage{booktabs}
\usepackage{xcolor}
\usepackage{authblk}
\usepackage{float}
\usepackage[authoryear]{natbib}
\usepackage[right=2.5cm,left=2.5cm,top=2cm,bottom=3cm]{geometry}
\usepackage[font={small,it}]{caption} 
\usepackage[colorlinks,linkcolor=blue,citecolor=blue,urlcolor=blue]{hyperref}
\usepackage{subfig}
\usepackage{mathtools}

\usepackage{multirow}
\usepackage{makecell} 
\usepackage{afterpage}

\newtheorem{assumption}{Assumption}
\newtheorem{proposition}{Proposition}

\usepackage{color}
\usepackage[normalem]{ulem}
\definecolor{orange}{RGB}{216,101,0}

\newcommand{\numreplicas}{100}

\title{ \textsc{Addressing Phase Discrepancies in Functional Data: A Bayesian Approach for Accurate Alignment and Smoothing} }
\author[1]{Jacopo Gardella}
\author[2]{Raffaele Argiento}
\author[3]{Alessandro Casa}
\author[4]{Alessia Pini}
\affil[1]{Department of Economics, Management and Statistics, Università degli Studi di Milano Bicocca}
\affil[2]{Department of Economics, Università degli Studi di Bergamo}
\affil[3]{Faculty of Economics and Management, Free University of Bozen-Bolzano}
\affil[4]{Department of Statistical Sciences, Università Cattolica del Sacro Cuore}
\date{}                     

\begin{document}
\maketitle

\begin{abstract}
In many real-world applications, functional data exhibit considerable variability in both amplitude and phase. This is especially true in biomechanical data such as the knee flexion angle dataset motivating our work, where timing differences across curves can obscure meaningful comparisons. Curves of this study also exhibit substantial variability from one another. These pronounced differences make the dataset particularly challenging to align properly without distorting or losing some of the individual curves characteristics. Our alignment model addresses these challenges by eliminating phase discrepancies while preserving the individual characteristics of each curve and avoiding distortion, thanks to its flexible smoothing component. Additionally, the model accommodates group structures through a dedicated parameter. By leveraging the Bayesian approach, the new prior on the warping parameters ensures that the resulting warping functions automatically satisfy all necessary validity conditions. We applied our model to the knee flexion dataset, demonstrating excellent performance in both smoothing and alignment, particularly in the presence of high inter-curve variability and complex group structures.
\end{abstract}

\section{Introduction}
\label{sec:introduction}
Curve alignment, also known as functional data alignment or warping, is a fundamental technique in the analysis of functional data. It involves adjusting the time or space domain of curves to ensure their key features, such as peaks, troughs, or other significant patterns, are properly aligned \citep{marron2014statistics,marron2015functional,vantini2012definition}. This process is crucial in many scientific and engineering disciplines, including biology, medicine, finance, and signal processing. Aligning curves enables researchers and analysts to achieve more accurate and insightful comparisons of data from various sources or subjects.
Mathematically, curve alignment is performed by applying a different monotone transformation function - often referred to as warping function - to the domain of each individual curve. 


\paragraph*{Knee flexion data}
The motivation for this work arises from a biomechanical application. In biomechanics, functional data are commonly used to analyze time-varying measurements - such as joint angles or muscle activity - to understand movement patterns and physiological functions. These measurements are typically collected during the execution of a specific task, such as a jump or a short walk. Because the duration of the task may vary across individuals, biomechanical data must be temporally registered to enable meaningful comparisons.

\begin{figure}[t]
    \centering
    \subfloat{
        \includegraphics[width=4.5cm]{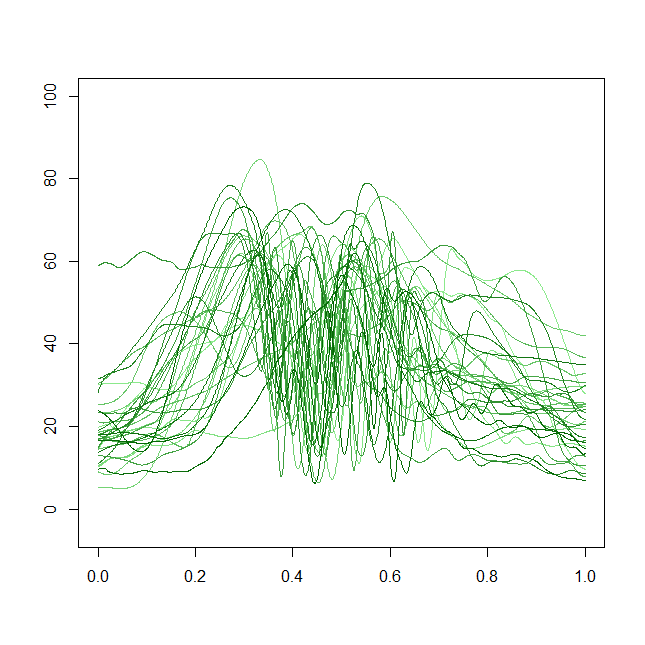}
    }
    \subfloat{
        \includegraphics[width=4.5cm]{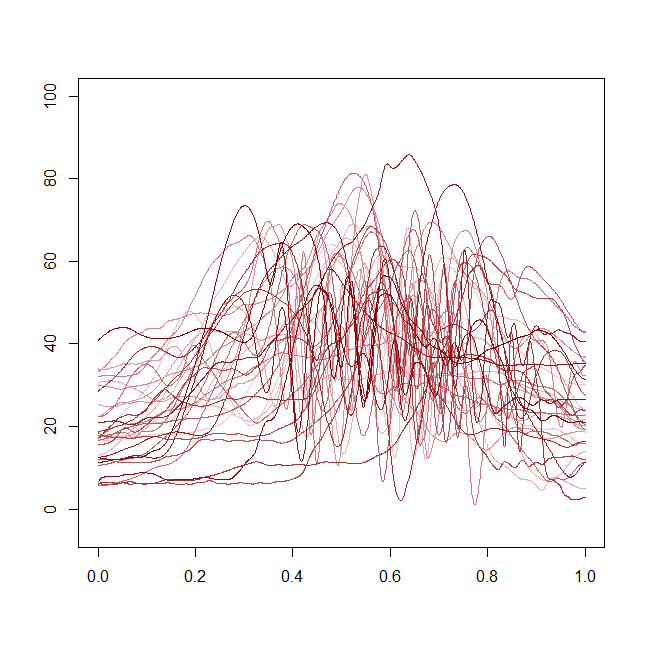}
    }
    \subfloat{
        \includegraphics[width=4.5cm]{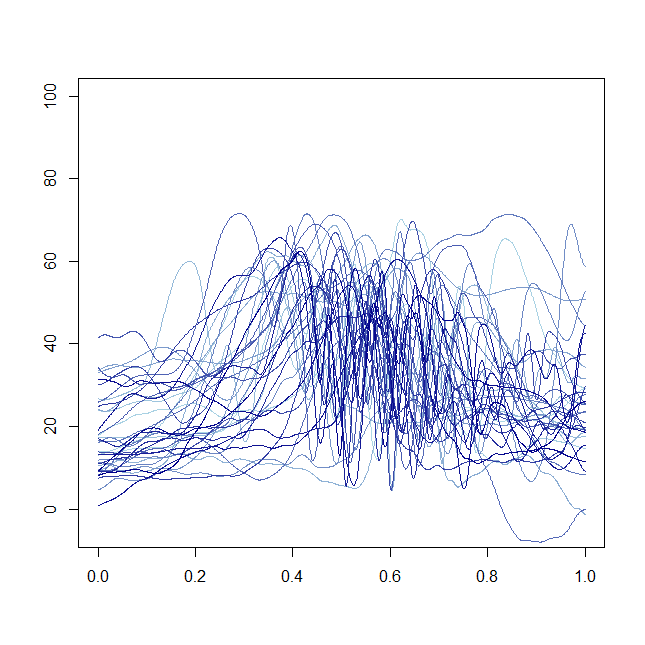}
    }
    \caption{Knee flexion angle as a function of the percentage of completion of the jump. Left: Control group; Center: Surgery group; Right: Physiotherapy group.}
    \label{fig:three_images}
\end{figure}

In this paper, we analyze a knee joint kinematic dataset from a follow-up study on subjects with anterior cruciate ligament (ACL) injuries \citep{tengman2015anterior}. The dataset consists of knee flexion angle measurements recorded during a one-leg hop for distance, across three groups: healthy controls, patients who received both surgery and physiotherapy (surgery group), and patients who received physiotherapy only (physiotherapy group). The flexion/extension angles were recorded over the full duration of the hop at a sampling rate of 240 Hz. Following standard biomechanical practice and consistent with prior analyses of this dataset \citep{tengman2015anterior}, measurements correspond to the non-dominant leg for the control group, and to the injured leg for the patient groups. The knee flexion angle of individuals of the three groups as a function of the percentage of completion of the jump is displayed in Figure~\ref{fig:three_images}.

The main purpose of this study is to align curves to compare the groups, and to understand the differences in kinematics between healthy subjects and the two injured patients' groups. A visual inspection of Figure \ref{fig:three_images} clearly highlights the unique challenges posed by these data structures from a statistical perspective. 
Indeed, data exhibit significant variability, even within the same group, with many curves showing unique individual characteristics. A large part of this variability is due to the fact that relevant features of the curves (e.g., peaks) occur at very different time points across individuals. Hence, in order to be able to compare groups, it is crucial to align or register data. 


An earlier study by \citet{Abramowicz2018} employed a landmark-based approach on a reduced version of the same dataset. To account for varying time lengths, the data were aligned using key features such as maximal knee flexion before takeoff, the takeoff and touchdown events, and maximal knee flexion after landing. However, this method have some potential drawbacks. For instance, the exact location of landmarks can be affected by noise, leading to a potential misalignment. Furthermore, the focus on single points during alignment may overlook discrepancies in other regions of the curves. Additionally, given the piecewise linearity of the warping functions, landmark registration technique artificially introduces discontinuities in the first derivatives of functional data. Finally, the method is not generalizable to the case where no landmark information is available. 

\paragraph*{Literature review}
Many  approaches have been introduced in the frequentist literature to overcome the above mentioned issues. For instance, \citet{Silverman} - later extended by \citet{MacGibbon,Li} - uses Functional Principal Component Analysis (FPCA) to estimate time-shift transformations. \citet{srivastava2010shape} propose to align curves using their square-root velocity function (SRVF). 
For a review and comparison of alignment techniques, the reader is referred to 
\citet{marron2014statistics,marron2015functional} and references therein. These methods primarily differ in the type of warping functions they allow (e.g., linear, affinities, or more complex functions). A common features of all these techniques is that alignment is treated as a preprocessing step to be performed before conducting inferential statistical analysis. Hence, the uncertainty inherent in the alignment process is not taken into account in the successive analyses, for instance when comparing the groups. 
In the nonparametric framework, there exist methods that perform simultaneously curve alignment and clustering \citep[see e.g.,][]{Liu,Sangalli}. \cite{Raket} developed a functional smoothing and registration model, addressing the dual challenge of reducing noise and aligning curves within a unified framework. However, none of these methods can perform a group comparison within the alignment setting. Given these limitations, a model-based Bayesian approach could offer a more robust solution by providing more accurate and comprehensive alignment results.

A Bayesian model for curve registration was proposed by \citet{Telesca}, leveraging the Bayesian framework to handle uncertainty, incorporate prior knowledge effectively, and accommodate complex model specifications, thus enhancing both inference robustness and alignment accuracy. Their model assumes that all curves share a common shape function, which can be shifted and scaled by individual-specific constants. The method is spline-based, as it models both the common shape and the warping functions using B-spline basis functions. 
Building on this work, \cite{Zachary} further extended the methodology, transforming it into a landmark-based approach.
\cite{Earls,Hooker} introduced a variation where each curve is represented as a weighted sum of two common curves, rather than a single one. This method increases flexibility,  allowing for a more nuanced representation of the data. Additionally, they replace the spline representation of the common terms with Gaussian processes, and replaced the MCMC inference method with an Adapted Variational Bayes algorithm.

Still within a Bayesian framework but adopting a distinct approach, \cite{Cheng} proposed an alignment model based on the quotient space. This method distinguishes the shape of functions from time-warping effects using the square root velocity function (SRVF), facilitating alignment based on shape while accommodating different timings. Similarly, the alignment procedures in the models introduced by \cite{Tucker} and \cite{Matuk} also use the SRVF, combined with Hamiltonian Monte Carlo (HMC) methods and Elastic Functional Data Analysis (EFDA), respectively. Additionally, \cite{Datta} proposed a Bayesian warping method for aligning fMRI activation maps. 

\paragraph*{Our proposal}
In this work, we base our alignment and smoothing model on the framework proposed by \citet{Telesca}. This choice is motivated primarily by the nature of our data: biomechanical functional data are typically recorded at high frequency and are intrinsically smooth, as abrupt movements are generally not possible in natural settings. As such, splines - known for their flexibility and effectiveness in modeling smooth functional data - are a particularly suitable modeling tool. Furthermore, the framework by \citet{Telesca} is readily extensible, enabling us to incorporate group-specific components without introducing undue complexity or compromising interpretability. While more recent approaches offer increased flexibility, they often come at the cost of greater model complexity or rely on assumptions that may not align as well with the smoothness characteristics of our data or the hierarchical structure we aim to capture. 

However, the model proposed by  \citet{Telesca} relies on a very simple smoothing component - a single common shape function - to facilitate easy alignment of the curves. Hence, it exhibits inherent limitations in flexibility: the common shape function can only be shifted or scaled for generating each individual curve, which may not adequately capture the unique variations across different subjects. The application of this warping technique to our dataset highlights important limitations, suggesting that this approach is not fully equipped to handle the complexity of such data. 
In contrast, it is possible to show that if a model with an overly advanced smoothing component is used to register the curves, the smoothed curves will be almost identical to the original ones, rendering the alignment process unnecessary since the curves can be fully described by the smoothing model. 

In this paper, building on the work of  \cite{Telesca}, we propose a novel Bayesian model for functional data alignment. Our approach is designed to accurately align datasets with highly variable curves, such as those in our specific application, ensuring effective handling of substantial differences between curves. 
For this reason, we propose a more complex model in terms of curve smoothing, which remains capable of accurately aligning the curves. Our model not only provides a more flexible smoothing method that better accommodates the individual characteristics of each curve but also includes a group-specific parameter. This extra parameter allows the model to automatically account for potential grouping structures when performing curve alignment, ensuring that it can simultaneously handle all the curves from different groups while incorporating group-specific information.

We also introduce a new framework for managing warping functions. In fact, these functions are the foundation of all registration models, as they are responsible for the actual alignment of the data. However, for these functions to be effective, they must satisfy certain assumptions, with their violation potentially inducing distortions during curve alignment. These requirements can be challenging to impose and may complicate the alignment process, potentially affecting the model's overall quality. Leveraging the Bayesian approach, we propose a new prior for the alignment parameters that inherently results in warping functions meeting all required conditions. 
 
This paper is structured as follows. 
First, we introduce our novel model in Section \ref{Model gen}. Its structure, detailed in Section \ref{Model structure}, consists of two main components: the smoothing part, covered in Section \ref{SMOOting model}, and the warping part, covered in Section \ref{warp model}, which includes our novel prior for the warping parameters, showcasing a key innovation of our approach.
Section \ref{Post_inf} describes the MCMC algorithm used in our analysis. It includes a detailed explanation of the full conditional distributions with closed-form solutions, as reported in Section \ref{Closed form}, and the Metropolis-Hastings steps for parameters lacking closed-form updates, as detailed in Section \ref{MH step our prior}.
In Section \ref{simualations}, we present the results of extensive simulation studies, where we test the performance of our model on two simulated datasets. We compare our results with those obtained using the method proposed by \cite{Telesca}, demonstrating the advantages and robustness of our approach.
We then present the results of our model on the knee flexion angle dataset in Section \ref{Resultsss}, providing concrete evidence of the effectiveness of our approach.
Finally, Section \ref{Final conclusion} offers concluding remarks and comments regarding the proposed model and the results achieved, summarizing the key findings and implications of our work.

\section{The Bayesian Hierarchical Model}
\label{Model gen}
We use a Bayesian hierarchical model for performing both smoothing and alignment of the curves. Parameters at the same hierarchical level are assumed to be a \emph{priori} independent. 

\subsection{Hierarchical Model for Groups of Curves}
\label{Model structure}
Let \( y_{gi}(t) \) denote the value of the \( i \)-th curve in group \( g \) at time \( t \), with \( i = 1, \ldots, N_g \), \(g = 1, \ldots, G \) and \( t \in [t_0, t_f] \), where $N_g$ is the number of individuals in the $g$-th group, $G$ is the number of groups and $t_0$,$t_f$ are the extreme points of the considered time domain. 
We assume that each individual has \( n_{gi} \) time observations, potentially varying among individuals, within the interval \( [t_0, t_f] \).
Then, we introduce the following sampling model for the curves:
\begin{align*}
  y_{gi}(t) & = \tilde{m}_{gi}(t) + \epsilon_{gi}(t)\, \nonumber,\\
  \tilde{m}_{gi}(t) &= m_{gi}(t) \circ h_{gi}(t) = m_{gi}(h_{gi}(t))\, ,\\
  m_{gi}(t) &= m_{gi}(t,\boldsymbol{\beta_g},\boldsymbol{\gamma}_{gi}) =  \mathcal{B}_{\beta}(t)^T \boldsymbol{\beta_g} + \mathcal{B}_{\gamma}(t)^T \boldsymbol{\gamma}_{gi}\, ,\\
   h_{gi}(t) &= \mathcal{B}_h(t)^T \boldsymbol{\phi}_{gi}\, ,\\
    \epsilon_{gi}(t) &\overset{iid}{\sim}\mathcal{N}(0, \sigma_{\epsilon}^2) \, ,  
\end{align*}
where $i=1,\dots,N_g$, $g=1,\dots,G$ and $ t \in [t_0,t_f]$. 

Therefore, we are assuming that the errors are homoscedastic, independent and identical distributed as a normal distribution with zero mean and $\sigma_{\epsilon}^2$ variance.
The term $m_{gi}(t)$ denotes the smoothing component of the model, which is responsible for reconstructing the curves. It consists of two components, with $\mathcal{B}_{\beta}(t)^T \boldsymbol{\beta_g}$ being shared by all the curves within the same group, while  $\mathcal{B}_{\gamma}(t)^T \boldsymbol{\gamma}_{gi}$ is specific to each individual curve. 
Further details on the smoothing component can be found in Section \ref{SMOOting model}. 
The term $ h_{gi}(t)$ represents the warping function associated with the individual $i$ in group $g$, evaluated at time $t$; readers are referred to Section \ref{warp model} for a detailed description. 

Finally, the registered curves are obtained by applying the inverse of the warping functions to the individual curves themselves that is
\begin{equation*}
      y_{gi}^*(t) = y_{gi}(t) \circ h_{gi}^{-1}(t)  = y_{gi}[h_{gi}^{-1}(t)]  \quad t \in [t_0,t_f] \, ,
\end{equation*}
for $g=1,\ldots,G,\ i=1,\ldots,N_g$.

 The Bayesian model is finalized by incorporating prior distributions for the model parameters, with parameters at the same hierarchical level treated as independent. At this stage, we specify the prior for the variance term $\sigma^2_\epsilon$, which is assigned a standard conjugate inverse-Gamma distribution. The priors for the remaining components are introduced in Sections \ref{SMOOting model} and \ref{warp model}.

\subsection{Smoothing Model}
\label{SMOOting model}
The modeling approach for the smoothing term $m_{gi}(t)$ is motivated by unreported exploratory analyses, highlighting a clear trade-off between the warping and the smoothing model. In fact, too flexible smoothing approaches such as the one by \citet{Dunson} tend to overfit the data, making subsequent data alignment procedures redundant. On the other hand, less flexible techniques such as the one considered by \citet{Telesca}, are overly rigid and seem to lack of the flexibility to capture relevant individual characteristics. To overcome these drawbacks, we propose an alternative smoothing model which, for the curve $i=1,\ldots,N_g$ in group $g=1,\ldots,G$ reads as
\begin{equation*}
     m_{gi}(t) = m_{gi}(t,\boldsymbol{\beta_g},\boldsymbol{\gamma}_{gi}) =  \mathcal{B}_{\beta}(t)^T \boldsymbol{\beta_g} + \mathcal{B}_{\gamma}(t)^T \boldsymbol{\gamma}_{gi}\,,    \quad t \in [t_0,t_f].
\end{equation*}

Therefore, our model incorporates two distinct components: a group-level effect, $\mathcal{B}_{\beta}(t)^T \boldsymbol{\beta_g}$, and an individual-level effect, $\mathcal{B}_{\gamma}(t)^T \boldsymbol{\gamma}_{gi}$. The former captures shared characteristics within a group, while the latter provides more flexibility in capturing the idiosyncratic features of each functional datum. To effectively model these complex functional forms, we utilize a cubic B-spline basis expansion, chosen for its inherent flexibility. In details,  the coefficients' vectors of the basis expansion are $\boldsymbol{\beta_g} \in \mathbb{R}^p$ for the common shape function of group $g$, and $\boldsymbol{\gamma_{gi}} \in \mathbb{R}^k$ for the individual shape function of the $i$-th patient in group $g$, while $\mathcal{B}_{\beta}(t)^T$ and $\mathcal{B}_{\gamma}(t)^T$ are respectively the $p$-dimensional and $k$-dimensional B-spline design vectors  evaluated at time $t$. Note that, as will be further discussed, we propose to use basis expansions for the two terms with different dimensions $p$ and $k$.

For the coefficients of the group-level effect $\boldsymbol{\beta_g}$, we consider a first-order random walk shrinkage prior, namely $\beta_{g_j} = \beta_{g_{j-1}} + e_j, e_j \overset{iid}{\sim}\mathcal{N}(0, \lambda)$, so that
\begin{align*}
    \boldsymbol{\beta_g} | \lambda &\overset{iid}{\sim} \mathcal{N}_p (\boldsymbol{0}, \boldsymbol{\Sigma_{\beta}}) \qquad g=1,...,G,\\
    \lambda &\sim \text{inv-gamma}(a_{\lambda}, b_{\lambda}) \, ,
\end{align*}
where $\boldsymbol{\Sigma_{\beta}}^{-1}=\frac{1}{\lambda}\boldsymbol{\Omega}$ 
is the tridiagonal precision matrix of a first-order random walk model, $\boldsymbol{\Omega}$ is a banded precision penalization matrix, 
and the random walk variance $\lambda$ can be interpreted as a smoothing parameter for the penalized regression splines \citep[see][for more details]{Telesca}.

For the individual specific parameters $\boldsymbol{\gamma_{gi}}$, we consider the following prior:
\begin{align*}
    \boldsymbol{\gamma}_{gi} | \sigma_{\gamma}^2 &\overset{iid}{\sim} \mathcal{N}_k(0, \boldsymbol{\Sigma_{\gamma}}) \qquad g=1,...,G \qquad i=1,...,N_g, \\
\boldsymbol{\Sigma_{\gamma}}&=\sigma_{\gamma}^2\mathbb{I}_k, \\
\label{prior_iniziale_gamma2}
    \sigma_{\gamma}^2 &\sim \text{inv-gamma}(a_{\gamma}, b_{\gamma}) \, ,
\end{align*}
where $\mathbb{I}_k$ is the identity matrix with dimension $k$. 
Differently from ${\boldsymbol{\beta}_g}$'s, we assume $\boldsymbol{\gamma}_{gi}$'s to be independent. This choice is driven by previous considerations on the trade-off between the smoothing and the warping components.

Moreover, $p$ has to be sufficiently greater than $k$. In fact, when $k \simeq p$ or even $k > p$, we encounter the issues mentioned above, leaving no room for proper alignment. Our recommendation is to treat the choices of \( p \) and \( k \) as specific to the dataset and application, and, in general, to select \( p \) to be moderately larger than \( k \). See Appendix  \ref{k almost p} for more details.

\subsection{Warping Model}
\label{warp model}
In this section, we describe the modeling approach used for the warping functions $h_{gi}(t)$. Since individual curves exhibit features occurring at different
times, we use a curve-specific random time-transformation
function for each individual in each group.

To qualify as a warping function defined over the set [$t_0$,$t_f$], a function must satisfy the following requirements:
\begin{assumption}\label{ass:assumption_A1}
    Time transformations must be strictly monotone increasing
\begin{equation*}
\label{Monotonicity hyp}
    h_{gi}(t_1) < h_{gi}(t_2) \quad \text{if}\;\; t_1 < t_2 \quad \text{for}\;\;i=1,\ldots,N_g, g=1,\ldots,G,\;\; \forall t_1,t_2 \in [t_0, t_f].
\end{equation*}
\end{assumption}
\begin{assumption}\label{ass:assumption_A2}
Time is constrained to be in the interval [$t_0$,$t_f$]: 
   \begin{equation*}
   \label{Image hyp}
         h_{gi}(t_0) = t_0 \qquad h_{gi}(t_f) =  t_f
        \qquad i=1,\ldots,N_g, \quad g=1,\ldots,G.
    \end{equation*}
    \end{assumption}
    \begin{assumption}\label{ass:assumption_A3}
    Continuity assumption 
    \begin{equation*}
    \label{Continuity hyp}
    \lim_{{t \to t_1}} h_{gi}(t) = h_{gi}(t_1) \qquad i=1,\ldots,N_g, \quad g=1,\ldots,G \qquad \forall t_1 \in [t_0, t_f].
\end{equation*}
\end{assumption}
Assumptions \ref{ass:assumption_A1} and \ref{ass:assumption_A2} are commonly applied to warping functions \citep[see][]{Ramsay}. These assumptions allow to maintain the ordering of observations along the domain and ensure that the domain remains unchanged before and after alignment. Additionally, Assumption \ref{ass:assumption_A3} guarantees that the registration process does not create gaps in time \citep[see][]{Zachary}.

Without loss of generality, the time interval [$t_0$,$t_f$] is mapped to $[0,1]$. In a more general setting with non-normalized time, the transformation
$\phi_{gij}' = t_0 + \phi_{gij}(t_f - t_0)  $, $j=1,...,q$, $ i=1,...,N_g $, $ g=1,...,G,$ can be used to obtain the vector's components in the interval $[t_0, t_f]$.

Coherently with Section \ref{SMOOting model}, warping functions are modeled as a linear combination of cubic B-spline basis functions with coefficients $\boldsymbol{\phi_{gi}}$. Also in this case,  
we define $\mathcal{B}_h(t)$ as the $q$-dimensional design vector of B-splines basis evaluated at time $t$. 
Summing up, we define the warping functions for curve $i=1,\ldots,N_g$ in group $g=1,\ldots,G$, as
\begin{equation*}
    h_{gi}(t) = \mathcal{B}_h(t)^T\boldsymbol{\phi_{gi}}, \qquad t\in [t_0, t_f] \, ,
\end{equation*}
where Assumptions \ref{ass:assumption_A1}, \ref{ass:assumption_A2} and \ref{ass:assumption_A3} translate into assumptions on the coefficients' vectors $\boldsymbol{\phi_{gi}}$.
In particular, Assumption \ref{ass:assumption_A3} is always satisfied by $h_{gi}(t) $ since cubic splines are twice differentiable, while Assumptions \ref{ass:assumption_A1} and \ref{ass:assumption_A2} require the following  conditions on the 
\emph{warping coefficients} parameters:
\begin{itemize}
    \item[C$_1$] To satisfy the strict monotonicity (increasing), as stated by \cite{DeBoor}, it suffices that 
    \begin{equation*}
        \phi_{gi1} < \phi_{gi2} <... < \phi_{giq}, \qquad i = 1,...,N_g, \qquad g=1,...,G.
    \end{equation*}
    This implies that the monotonicity of the components of $\boldsymbol{\phi_{gi}}$ is sufficient for the monotonicity of the warping function $h_{gi}(t)$.
    \item[C$_2$] To satisfy the image constraint, it is sufficient to impose   $\phi_{gi1}=t_0$, $\phi_{giq}=t_f$, for $i = 1,...,N_g$ and $g=1,...,G$.
\end{itemize}
In this work, we introduce a new prior distribution for $\boldsymbol{\phi_{gi}}$ ensuring that conditions C$_1$ and C$_2$ are satisfied both \emph{a priori} and \emph{a posteriori}, without any need for transformation or adjustment of these parameters.
In detail, the joint distribution of the constrained coefficients $\boldsymbol{\phi_{gi}}$ is given intrinsically by the following construction. 
First, independently for each $j=2,...,q;$ $i=1,...,N_g;$ $g=1,...,G$ we introduce a latent random variable $\xi_{gij}\sim \text{gamma}(a_j,b)$, then the coefficients are built by normalization: 
\begin{equation*}
\phi_{gij} =   \frac{\sum_{k=1}^{j} \xi_{gik}}{T_{gi}} \, ,
\end{equation*}
where $\xi_{gi1} =0, T_{gi} = \sum_{k=1}^{q}  \xi_{gik}$.
The prior distribution ensures C$_1$ and C$_2$, thereby guaranteeing Assumptions \ref{ass:assumption_A1} and \ref{ass:assumption_A2} for our model's warping functions $h_{gi}(t)$.

Interestingly, the marginal distribution of the coefficient can be computed as stated in the following result (see Appendix \ref{Beta proof} for the proof).
\begin{proposition}
\label{prop:marginal}
For each $i=1,...,N_g$ and $g=1,...,G$, the 
joint distribution of the vector of  $\boldsymbol{\phi}_{gi}$ 
is characterized as follows:
\begin{enumerate}
    \item The joint law of the vector of the increments is a Dirichlet distribution, that is 
    \begin{equation*}
        (\phi_{gi,2}, \phi_{gi,3} -\phi_{gi,2},...,1-\phi_{gi,q-1}) \sim \text{Dir}(a_2,...,a_q).
    \end{equation*}
    
    \item $T_{gi}$ follows a gamma distribution and is independent from $\boldsymbol{\phi}_{gi}$.
 \item The marginal distribution of each component is given by:
    \begin{equation*}
    \label{eq:marginal dist of phi_gij}
        \phi_{gi,j} \sim
    \begin{cases}
        \delta_0  & \text{if } j = 1, \\
        \text{Beta}(\alpha_{1j}, \alpha_{2j}) & \text{if } 2 \leq j \leq q-1, \\
        \delta_1 & \text{if } j = q.
    \end{cases}
    \end{equation*}
    where \(\delta_s\) refers to a Dirac delta distribution centered at \(s\) and 
    \begin{align*}
     \alpha_{1j}  &= \sum_{k=2}^{j} a_k \qquad\hspace{0.35cm} j=2,\dots,q-1,\\
     \label{Beta2 part 2}
        \alpha_{2j} &= \sum_{k=j+1}^{q} a_k \qquad j=2,\dots,q-1.
\end{align*}

\end{enumerate}
\end{proposition}

From Point 3 of Proposition \ref{prop:marginal}, it is straightforward to compute the expected value of $\phi_{gij}$ for $j=1,...,q, i=1,...,N_g, g=1,...,G$ that is
\begin{equation*}
    \mathbb{E}[\phi_{gij}] =
    \begin{cases}
        0  & \text{if } j = 1, \\
        \displaystyle\frac{\sum_{k=2}^{j} a_k}{\sum_{k=2}^{q} a_k} &  \text{if } 2 \leq j \leq q-1, \\
        1 & \text{if } j = q.
    \end{cases}
\end{equation*}

The result in Proposition \ref{prop:marginal} essentially states that $\boldsymbol{\phi}_{gi}$ is the cumulative sum of a Dirichlet random vector. From this latter observation (see Appendix \ref{app:covariance} for details), we obtain that
for $l \le m$ and $l, m = 2,..,q-1$
\begin{equation*}
    \text{Cov}(\phi_{gil},\phi_{gim})= \frac{\sum_{j=2}^l a_{gij} \sum_{k=m+1}^q a_{gik}}{(\sum_{s=2}^{q}a_{gis} +1) (\sum_{s=2}^{q}a_{gis})^2}.
\end{equation*}

The availability of a closed expression for the mean and the covariance of $\boldsymbol{\phi}_{gil}$ allows for prior elicitation and could be used, for instance, to incorporate in the model information about landmarks, as discussed in Section \ref{Final conclusion}.


\section{Posterior Inference}
\label{Post_inf}

We perform posterior inference using a Markov Chain Monte Carlo (MCMC) algorithm.
Our prior choice ensures that all full conditionals are available in closed form, except for those of the warping parameters $\boldsymbol{\phi_{gi}}$'s.
To update the latter, we implemented a Metropolis-Hastings (MH) step. The full conditionals are reported here. Detailed calculations for all parameters with a closed-form full conditional distribution can be found in Appendix \ref{Full cond count}, while Appendix \ref{MH appendix} provides further information on the MH step for the warping parameters.

\subsection{Joint Posterior Distribution}
To simplify notation, we assume that all $y_{gi}$ have the same number of observations, i.e. $n_{gi} = n$ for all $i=1,\dots,N_g$ and $g=1,\dots,G$. Note that generalization to the case where the number of observations differs is straightforward.  
Let $Y = (y_{11}$,\dots,$y_{1N_1}$,\dots,$y_{GN_G})$ be the full data matrix where each column is given by ${y}_{gi} \in \mathbb{R}^{n}$, the vector of observations of the $i$-th individual in group $g$.
Then let define the matrices $\Phi$ = ($\phi_{11}$,...,$\phi_{1N_1}$,...,$\phi_{GN_G}$), B = ($\boldsymbol{\beta_1}$,...,$\boldsymbol{\beta_G}$), $\Gamma$ = $(\gamma_{11}, \ldots, \gamma_{1N_1}, \ldots, \gamma_{GN_G})$, $\Xi=$$(\xi_{11}, \ldots, \xi_{1N_1}, \ldots, \xi_{GN_G})$. 
The posterior distribution is given by: 
\begin{align*}
 f(B, \Gamma, \Phi, \Xi, \lambda, \sigma_{\gamma}^2, \sigma_{\epsilon}^2 \mid Y )  &\propto f(Y \mid B,\Gamma,\Phi,\sigma_{\epsilon}^2) \, f(B \mid \lambda) \, f(\lambda \mid a_{\lambda},b_{\lambda}) \\
    &\quad \times f(\Gamma \mid \sigma_{\gamma}^2) \, f(\sigma_{\gamma}^2 \mid a_{\gamma},b_{\gamma}) \, f(\Phi \mid \Xi) \\
    &\quad \times f(\Xi \mid a_2,\ldots,a_q,b) \, f(\sigma_{\epsilon}^2 \mid a_{\epsilon},b_{\epsilon}).
\end{align*}

\subsection{Full Conditionals in Closed Form}
\label{Closed form}
In this subsection, we list all the available closed-form full conditionals (see Appendix \ref{Full cond count} for their derivation).
To ease notation, we report here the range of the indices: $i=1,\dots, N_g$ and $g=1,\dots, G$.
\begin{itemize}
\item Group-specific shape parameters  $\boldsymbol{\beta_g}$:
\begin{align*}
\pi(\boldsymbol{\beta_g}| -) & \sim \mathcal{N}_p (\boldsymbol{m_{\beta_g}}, \boldsymbol{V_{\beta_g}}),\\
\boldsymbol{V^{-1}_{\beta_g} } & = [ \boldsymbol{\Sigma^{-1}_{\beta}} + \frac{1}{\sigma_{\epsilon}^2} \boldsymbol{X_g^{T}}\boldsymbol{X_g}],\\
\boldsymbol{m_{\beta_g} } & = \boldsymbol{V_{\beta_g} }[ \frac{1}{\sigma_{\epsilon}^2} \boldsymbol{X_g^{T}}(\boldsymbol{Y_g} -\boldsymbol{C_{\gamma_g}})],\\ 
\boldsymbol{C_{\gamma_g}} & = (\mathcal{B}_{\beta}^T(h_{g1}(t_{g1},\phi_{g1}))\gamma_{g1},...,\mathcal{B}_{\beta}^T(h_{gN_g}(t_{gN_g},\phi_{gN_g}))\gamma_{gN_g}),\\
\boldsymbol{X_g} & = (\mathcal{B}_{\beta}^T(h_{g1}(t_{g1},\phi_{g1})),...,\mathcal{B}_{\beta}^T(h_{gN_g}(t_{gN_g},\phi_{gN_g}))).
\end{align*}
\item Common shape variance parameter $\boldsymbol{\lambda}$:
\begin{align*}
\pi(\lambda| -) & \sim  inv-gamma (a_{\lambda}^*,b_{\lambda}^*),\\
a_{\lambda}^{*} = a_{\lambda} +  \frac{Gp}{2}, & \qquad
b_{\lambda}^{*} = b_{\lambda} +  \frac{1}{2}\sum_{g=1}^{G} \boldsymbol{\beta_g}^T \Omega \boldsymbol{\beta_g}.
\end{align*}
\item 
Individual-specific shape parameters $\boldsymbol{\gamma_{gi}}$:
\begin{align*}
\pi(\boldsymbol{\gamma_{gi}}| -) &\overset{\text{ind}}{\sim} \mathcal{N}_k (\boldsymbol{m_{\gamma_{gi}}}, \boldsymbol{V_{\gamma_{gi}}}),\\
    \boldsymbol{V^{-1}_{\gamma_{gi}} } & = \left[ \boldsymbol{\Sigma^{-1}_{\gamma}} + \frac{1}{\sigma_{\epsilon}^2} \mathcal{B}_{\gamma}(h_{gi}(\boldsymbol{t_{gi}},\boldsymbol{\phi_{gi}})) \mathcal{B}_{\gamma}^T(h_{gi}(\boldsymbol{t_{gi}},\boldsymbol{\phi_{gi}})) \right],\\
    \boldsymbol{m_{\gamma_{gi}} } & = \boldsymbol{V_{\gamma_{gi}} } \left[ \frac{1}{\sigma_{\epsilon}^2} \mathcal{B}_{\beta}(h_{gi}(\boldsymbol{t_{gi}},\boldsymbol{\phi_{gi}}))(\boldsymbol{Y_{gi}} - \boldsymbol{C_{\beta_g}^{gi}}) \right],\\
    \boldsymbol{C_{\beta_g}^{gi}} & = \mathcal{B}_{\beta}^T(h_{gi}(\boldsymbol{t_{gi}},\boldsymbol{\phi_{gi}}))\boldsymbol{\beta_g}.
\end{align*}
\item 
Variance $\sigma^2_{\gamma} $ of the individual specific vectors $\boldsymbol{\gamma_{gi}}$, i=1,...,$N_g$, g=1,...,G: 
\begin{align*}
    \pi(\sigma^2_{\gamma} | -) &\sim inv-gamma(a_{\gamma}^*,b_{\gamma}^*), \\
    a_{\gamma}^*=a_{\gamma} +  \frac{Nk}{2}, & \qquad
    b_{\gamma}^*= b_{\gamma} +  \frac{1}{2}\sum_{g=1}^{G} \sum_{i=1}^{N_g}\boldsymbol{\gamma_{gi}}^T\boldsymbol{\gamma_{gi}}.
\end{align*}
\item 
Error variance parameter
$\sigma^2_{\epsilon}$:
\begin{align*}
    \pi(\sigma^2_{\epsilon}| -) &\sim inv-gamma(a_{\epsilon}^*,b_{\epsilon}^*),\\
    a_{\epsilon}^*&=a_{\epsilon} +  \frac{1}{2}\sum_{g=1}^{G} \sum_{i=1}^{N_g} n_{gi},\\
    b_{\epsilon}^*&= b_{\epsilon} +  \frac{1}{2}\sum_{g=1}^{G} \sum_{i=1}^{N_g} (\boldsymbol{Y}_{gi} - \boldsymbol{\tilde{m}}_{gi})^{T} (\boldsymbol{Y}_{gi} - \boldsymbol{\tilde{m}}_{gi}),\\
   \boldsymbol{\tilde{m}}_{gi}&=\mathcal{B}_{\beta}(h_{gi}(\boldsymbol{t_{gi}},\boldsymbol{\phi_{gi}}))^T \boldsymbol{\beta_g} + \mathcal{B}_{\gamma}(h_{gi}(\boldsymbol{t_{gi}},\boldsymbol{\phi_{gi}}))^T \boldsymbol{\gamma}_{gi} .
\end{align*}
\end{itemize}

\subsection{Full Conditional of the Warping Parameters}
\label{MH step our prior}
As previously mentioned, the full conditional distribution of $\boldsymbol{\phi}_{gi}$ is not available in closed form. 
The full conditional distribution of $\boldsymbol{\phi}_{gi}$ is not only analytically intractable, as previously mentioned, but also challenging to sample from, given the intricate dependency of the B-spline basis functions on $\boldsymbol{\phi}_{gi}$.''
Consequently, we use a Metropolis-Hastings step to sample the warping parameters.
Given the \emph{a priori} independence of the parameters $\boldsymbol{\phi}_{gi}$, and the fact that the likelihood factorizes over $g = 1, \dots, G$ and $i = 1, \dots, N_g$, it follows that the $\boldsymbol{\phi}_{gi}$ remain conditionally independent \emph{a posteriori}, given all other parameters. As a result, the update can be performed independently for each individual within each group. This structure reduces computational complexity and enables more efficient sampling -- for instance, through parallel updates. 
Moreover, the sampling procedure is simplified by defining $\boldsymbol{\phi}_{gi}$ through the normalization of $\boldsymbol{\xi}_{gi}$. The vector $\boldsymbol{\phi}_{gi}$ must satisfy Conditions \ref{ass:assumption_A1}–\ref{ass:assumption_A3} (see Section \ref{warp model}). Rather than enforcing these conditions directly on $\boldsymbol{\phi}_{gi}$, we sample from the full conditional distribution of $\boldsymbol{\xi}_{gi}$ and rely on the normalization procedure described in Section \ref{warp model} to ensure that the resulting $\boldsymbol{\phi}_{gi}$ automatically satisfies the required conditions \emph{a posteriori}.

As a final point, we mention that an Adaptive Scaling within the Adaptive Metropolis–Hastings algorithm \citep{griffin2013advances}  has been implemented to sample from the full conditional distribution of the unnormalized parameters $\boldsymbol{\xi}_{gi}$.

\section{Simulation study}
\label{simualations}
In this section, we evaluate the performance of the proposed model using simulated datasets. This allows to assess the model's performance in aligning the data, while preserving essential individual characteristics and avoiding distortion of important features.  We also compare our proposal with the model proposed by \cite{Telesca}. Performance is evaluated using the average $L_2$ distance between the original and the aligned curves. All the presented results are obtained using R code available at \cite{githubrepo}.
We test the two models on two different settings. In the first one, we simulate data from $G = 2$ groups, each composed of $N_g = 10$ ($g=1,2$) curves. All curves have the same number of observations, $n_{gi} = 300$. To model the shape of these curves, we used the smoothing model described in Section \ref{SMOOting model} with dimensions $p=13$, $k=7$, $q=10$ and parameters $\lambda = 5$, $\sigma_{\gamma}^2=5$ and $\sigma_{\epsilon}^2 = 0.01$. In the second setting, we simulate data from $G = 1$ group with $N_1 = 30$ curves, each with $n_{1i} = 300$ observations. The shape of these curves is obtained using the smoothing model proposed by \cite{Telesca} (see Appendix \ref{Appenddix:Telesca simulated data} for the details). In both settings, misalignment is induced using vectors $\boldsymbol{\phi}_{gi}$ obtained under the following model:
\begin{align*}
    \boldsymbol{\phi}_{gi} & = \frac{\sum_{k=1}^{j} \xi_{gik}}{\sum_{k=1}^{q} \xi_{gik}}, \qquad i=1,...,N_g, \qquad g=1,...,G,\\
    \xi_{gij}&\sim gamma(a_{gij},b_{gij}), \qquad j=2,...,q, \qquad i=1,...,N_g, \qquad g=1,...,G,\\
    \xi_{gi1}& = 0 \qquad i=1,...,N_g, \qquad g=1,...,G \, ,
\end{align*}
where the parameters $a_{gij}$ and $b_{gij}$ are  -- for each individual in every group -- randomly sampled on a grid. 

\begin{figure}[H]
    \centering
    \subfloat{
        \includegraphics[width=0.45\textwidth]{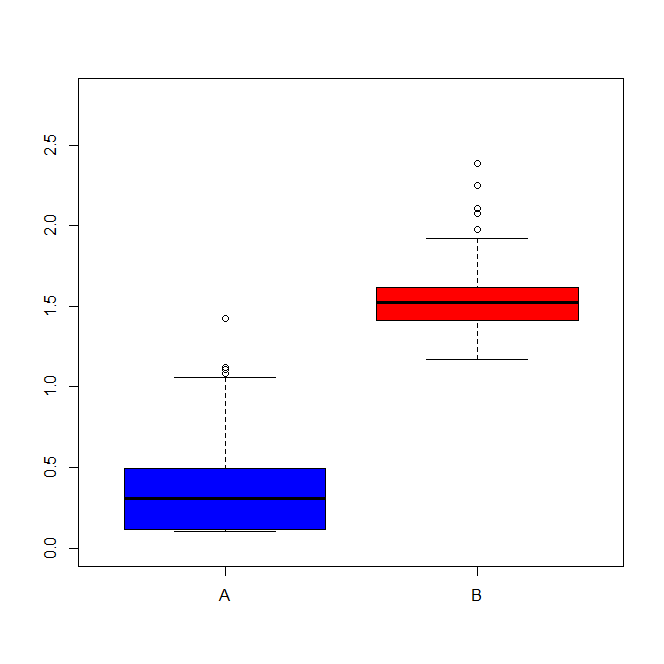}
    }
    \quad
    \subfloat{
        \includegraphics[width=0.45\textwidth]{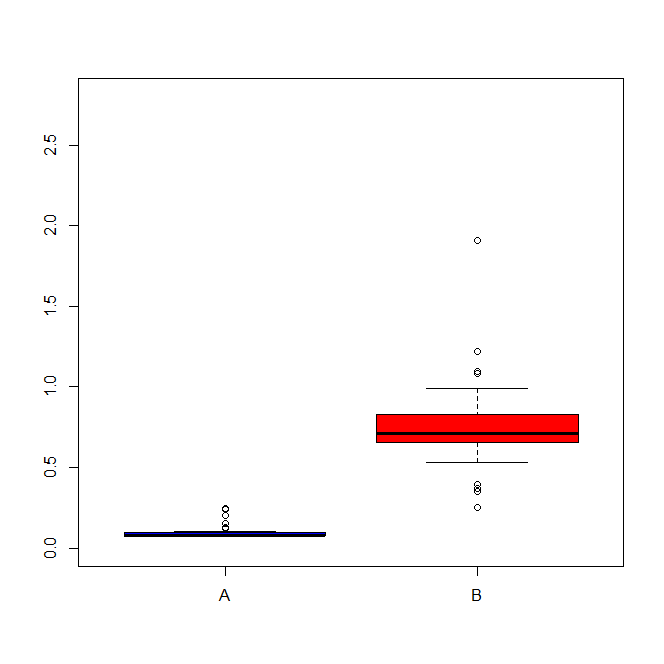}
    }
    \caption{Left: error comparison for the first setting; right:  Error comparison for the second setting. $A$ stands for the proposed model while $B$ indicates the model proposed by \citet{Telesca}.
    }
    \label{fig:MSE simulated data}
\end{figure}

\begin{figure}[H]
    \centering
    \subfloat{
        \includegraphics[width=0.45\textwidth]{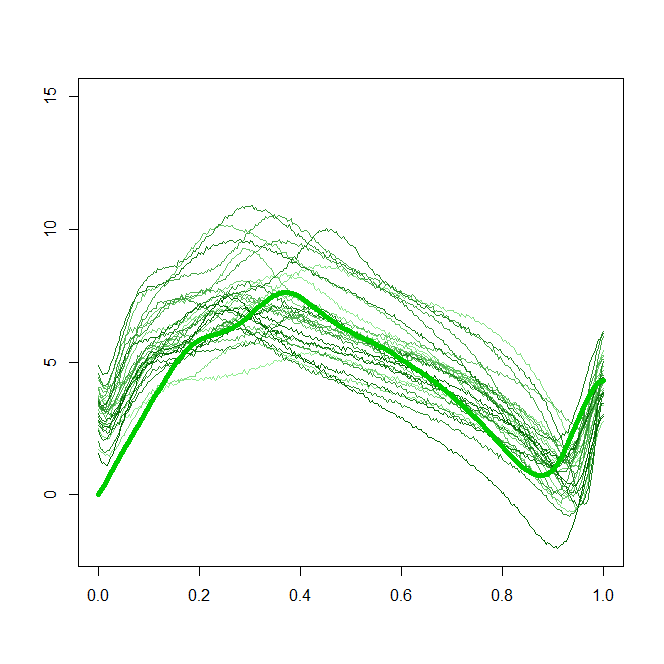}
    }
    \quad
    \subfloat{
        \includegraphics[width=0.45\textwidth]{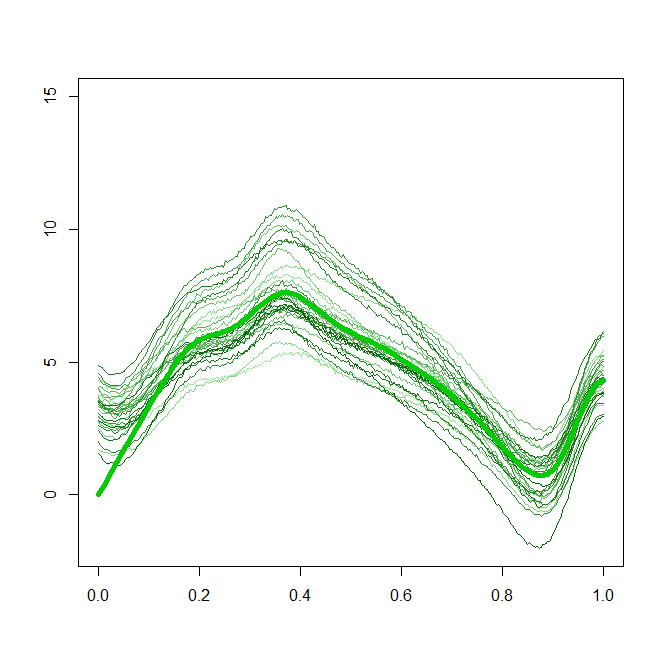}
    }
    \caption{Left: misaligned curves of one simulated dataset in setting 2; right: aligned curves using our proposed model. The thickest green curve represent the estimated common term.
    }
    \label{fig:aligned simulated curves}
\end{figure}

Results are obtained from $\numreplicas$ replicas of each model on each setting. Note that, since the model by \citet{Telesca} does not have a group-specific component, we applied it separately to the two groups, averaging the respective errors. This allows for a fair comparison between the two approaches even in the presence of distinct groups. 
Figure \ref{fig:MSE simulated data}  reports the $L_2$ distance between the true and the aligned curves obtained from the respective models.  As expected, both models exhibit a higher error on setting 1, which is more complex, while the error is lower on setting 2, where the curves have simpler and more similar shapes. Instead fixing the same setting, we observe that our model performs better, with significantly lower errors compared to the model proposed by \cite{Telesca}. This demonstrates the effectiveness of the proposed model and its great flexibility in aligning the curves while preserving their individual characteristics. While in setting 1 this could be partly due to the increased sample size (the model by \cite{Telesca} has been applied separately to the two groups), the difference is also evident in setting 2, where both models are applied to the same data set.
As an example, Figure \ref{fig:aligned simulated curves} shows the misaligned curves of an instance of simulated data from setting 2 (on the left) and the aligned curves obtained using our model (on the right). 
The common term (thickest curve) starts from zero because we chose to include the intercept only in the individual specific terms.
Our model exhibits strong performance compared to the competing approach, even when applied to a single group as in setting 2. In fact, beyond incorporating a hierarchical structure, one of its key advantages lies in the flexibility introduced by modeling the individual-specific term using a set of splines, rather than adopting a linear transformation as in \cite{Telesca}.

\section{Application}
\label{Resultsss}
In this section, we present the results of our model applied to the knee flexion angle dataset.  
The dataset consists of curves representing the knee flexion angle of 100 individuals during a complete one-leg hop. The objective of the study is to evaluate the long-term consequences of anterior cruciate ligament (ACL) rupture and its treatment method on individuals' movement patterns.
The individuals belong to three groups: 
\begin{itemize}
    \item Control group: 32 healthy individuals.
    \item Surgery group: 32 ACL-injured patients who have undergone surgery and physiotherapy.
    \item Physiotherapy group: 36 ACL-injured patients who have undergone physiotherapy only.
\end{itemize}
The number of observations for each curve varies significantly, even within the same group, ranging from a minimum of 303 to a maximum of 591. Through an initial exploratory analysis, we observe significant variability between the curves within each group, and this holds for all three groups. In particular, beyond the evident phase variability, which we aim to significantly reduce with our proposed method, there is also substantial amplitude variability. In particular, the dataset contains curves with different shapes and a varying number of peaks from one another. This strong variability makes it challenging to align them using models based on landmarks \citep[see e.g.][]{Abramowicz2018} without altering or losing the individual characteristics of each curve. For this reason, the greater flexibility of our model holds the potential to achieve better results on this specific dataset.

The results are obtained with $25000$ MCMC iterations, with the first $20000$ iterations discarded as burn-in. Convergence was assessed using standard diagnostic tools. We set the hyperparameters to \( a_{\epsilon} = 3000 \), \( b_{\epsilon} = 5000 \), \( a_{\lambda} = 1200 \), \( b_{\lambda} = 3500 \), \( a_{\gamma} = 1000 \), \( b_{\gamma} = 2000 \), and \( b = 2.5 \), aiming to prevent convergence issues by avoiding excessively large values while still allowing adequate exploration of the parameter space. Robustness checks were performed by varying the hyperparameter values, confirming that the results remained stable across a reasonable range of settings. Using the result in Proposition~\ref{prop:marginal}, we set the warping hyperparameters \( a_2, \dots, a_q \) so that the prior mean of the warping function corresponds to the identity, i.e., \( \mathbb{E}(h_{gi}(t)) = t \), and to avoid numerical instability issues.
The dimensions of the vector parameters were set to \( p = 23 \), \( k = 8 \), and \( q = 10 \) to ensure that all the significant information in the curves is captured and that proper alignment is achieved.

\begin{figure}[H]
    \centering
    \subfloat{
        \includegraphics[width=4cm]{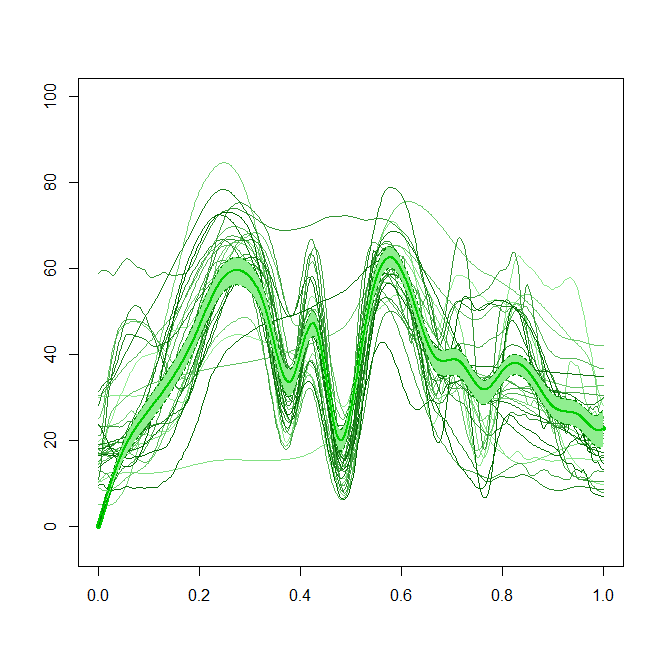}}
    \quad
    \subfloat{
        \includegraphics[width=4cm]{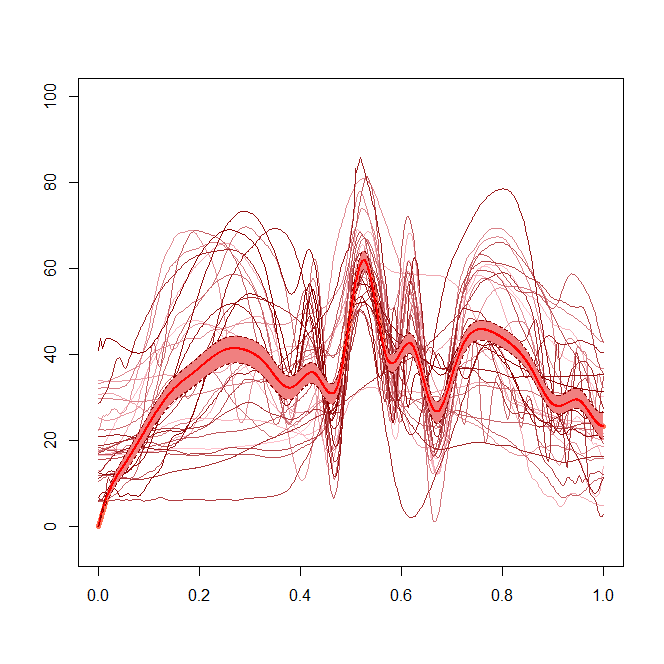}
    }
    \quad
    \subfloat{
        \includegraphics[width=4cm]{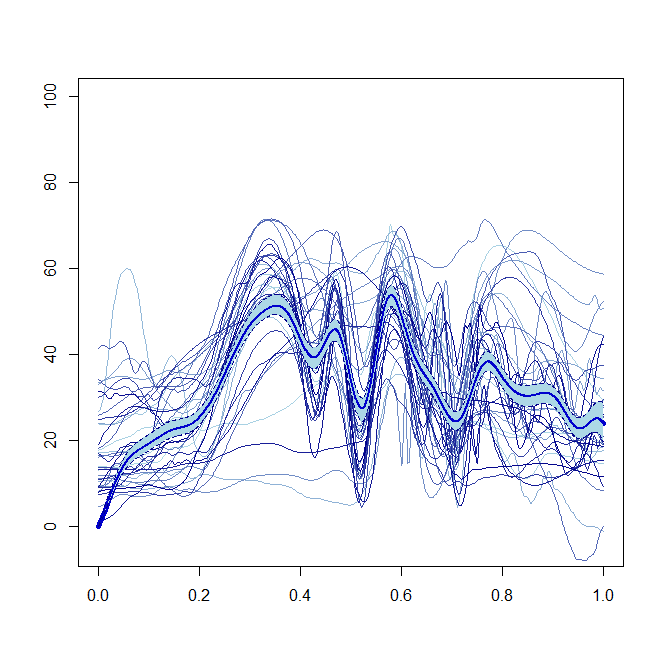}
    }
    \caption{Aligned curves divided by groups. Left: control group; center: surgery group; right: Physiotherapy group. The three thickest curves represent the estimated common term for each group together with their 95\% credibility bands: green for control, red for surgery, and blue for physiotherapy.}
    \label{fig:Knee data aligned}
\end{figure}

\begin{figure}[H]
    \centering
    \subfloat{
    \includegraphics[width=4cm]{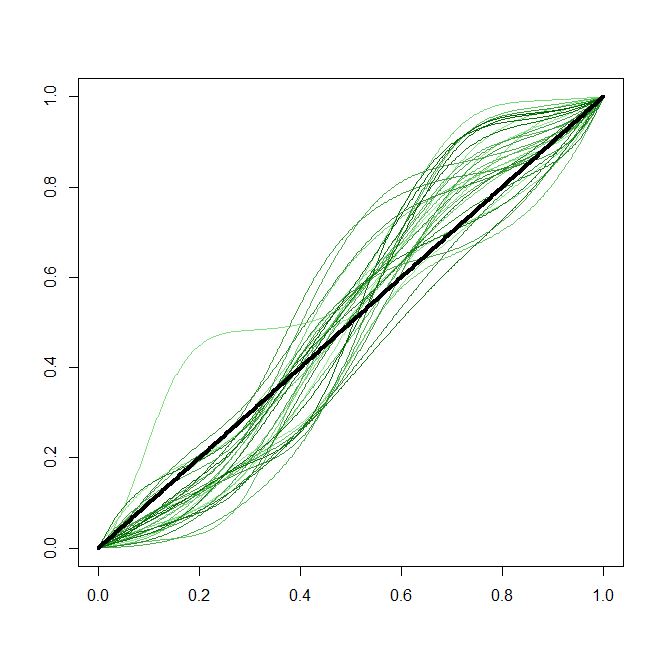}
    }
    \quad 
    \subfloat{
    \includegraphics[width=4cm]{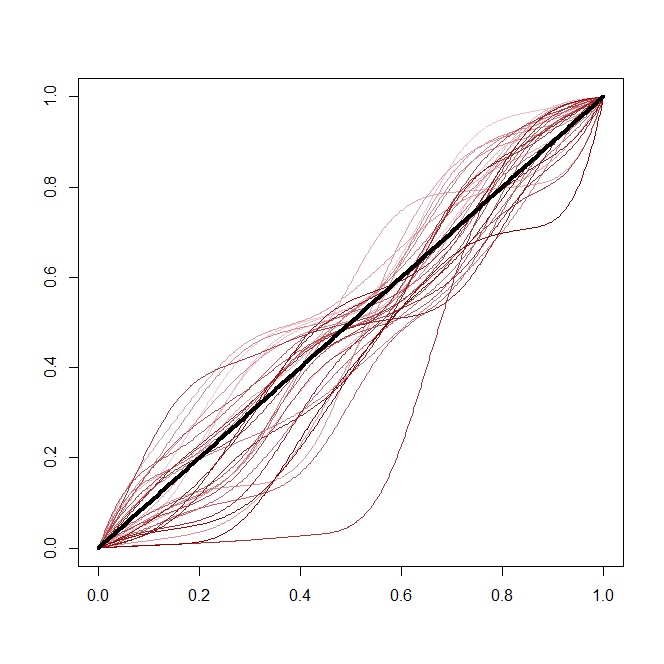}
    }
    \quad
    \subfloat{\includegraphics[width=4cm]{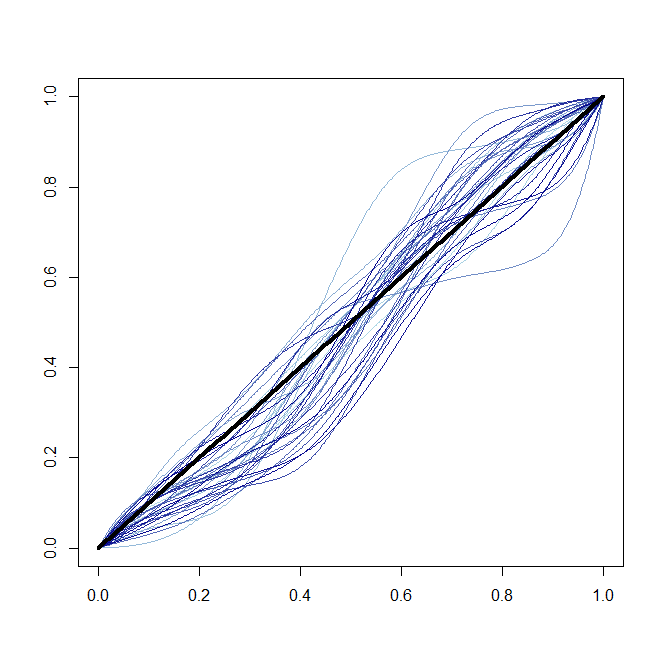}
    }
    \caption{Group-specific warping functions. Left: control group; center: surgery group; right: physiotherapy group. The three thickest curves represent identity line (i.e., no warping).}
    \label{fig:knee warping functions}
\end{figure}

The aligned curves are shown in Figure \ref{fig:Knee data aligned}, and the corresponding estimated warping functions in Figure \ref{fig:knee warping functions}.
The aligned curves exhibit a significant  reduction in phase variability, indicating that the model has successfully captured the underlying heterogeneity.
We can observe that both the estimated common terms and the alignment tend to be similar across the three different groups, thanks to the information sharing provided by the model. Examination of the aligned curves, both at the group and individual levels, reveals that the model achieves alignment while preserving the distinctive features of each curve, with no evident deformation or information loss.
The common trend often displays more peaks than individual curves; in such cases, the model achieves alignment by mapping the fewer peaks of the individual curves to a relevant subset of those in the common trend. Conversely, curves with a higher number of peaks are aligned across the entire sequence of peaks in the trend/mean curve. We point out that the estimated common curves pass through zero at time zero because the model is specified in such a way that the behavior of the curve near the origin is captured by the individual specific term $ \mathcal{B}_{\gamma}(t)^T \boldsymbol{\gamma}_{gi}$ as shown in the Appendix \ref{sec:intercept individual term}.
Summing up, the model is capable of aligning curves that exhibit more peaks, as well as those that have fewer. We observe that curves from the surgery and physiotherapy groups exhibit more variability and a more complex common shape compared to those from the control group. This is in line with previous analyses on the same data, which highlighted a significant long-term impact of ACL rupture on movement patterns \citep{Abramowicz2018}. In particular, the common component of the physiotherapy group exhibits the highest number of peaks, consistently with reports of increased knee instability in this group \citep{stensdotter2013deficits}. This finding is expected, since the individuals in the physiotherapy group no longer have an intact ACL, as it was not surgically reconstructed after the rupture.
Finally, we applied the \cite{Telesca} model to align the same curves, to provide a comparison between the two models on real data as well. The result is reported in the Appendix \ref{Appendix: Telesca alignment}. 
The model reduces phase variability effectively; however, some curves appear compressed or distorted to conform to the estimated common trend. As previously noted, the warping function in this competing model is somewhat rigid compared to our approach. As a result, it struggles to align highly diverse curves while preserving their individual characteristics.

Figure~\ref{fig:knee warping functions} shows that the warping functions for the Control and Physiotherapy groups exhibit similar behavior. In both cases, during the initial interval (approximately from 0 to 0.6), the functions satisfy \( t < h(t) \), while in the interval from 0.6 to 1, the relationship reverses to \( t > h(t) \). In contrast, the Surgery group displays a much more heterogeneous pattern. Interestingly, the point around \( t = 0.6 \) consistently marks a change in the pacing of the curves across all groups.

\section{Conclusions/Discussion}
\label{Final conclusion}
In this paper, we proposed a flexible Bayesian hierarchical model to jointly perform smoothing and functional registration, providing a natural framework for assessing uncertainty in the estimated time-transformation. 

Our model builds upon the baseline proposed by \cite{Telesca}, introducing a significantly more flexible smoothing component by incorporating an entire individual-specific curve alongside the group-specific trend, rather than relying solely on a multiplicative constant and a vertical shift. Inspired by the smoothing framework of \cite{Kowal}, we remove the assumption of a universal common curve, allowing instead for a distinct trend within each group - an advantage when dealing with heterogeneous data structures as in the knee flexion dataset considered in this work. Nonetheless, the model remains applicable in the absence of grouping information, defaulting to a structure where each individual has its own smoothing component and shares a common global curve, as detailed in Appendix \ref{No groups}. 
The hierarchical structure of our model introduces an additional gain by allowing for the sharing of information across groups of data. This proved particularly beneficial in the knee flexion data application in Section \ref{Resultsss}.
Notably, by leveraging the Bayesian approach, we have proposed a new prior for the warping parameters, which consistently results in valid warping functions without the need for additional transformations, thereby enhancing the robustness of the procedure. 


Our model demonstrates strong performance across a variety of scenarios. It not only outperforms state-of-the-art methods on complex, irregular curves -such as those in the knee flexion angle dataset - but also achieves superior results on more homogeneous shapes, as shown in the simulation study in Section \ref{simualations}. The proposed approach ensures accurate alignment without distorting or compressing the shape of individual curves. The flexible smoothing component plays a crucial role, allowing most curves to retain their original shape. As a result, the common component becomes more complex, effectively adapting to the variability across curves.

Leveraging the trade-off between alignment and smoothing discussed in Section \ref{SMOOting model}, future work could involve a deeper exploration of alternative smoothing methods, from simpler approaches, such as kernel smoothing and local polynomial smoothing \citep[see, e.g.,][for more detailed reviews]{Ramsay,ZhangBook}, to more complex ones such as the model proposed by \citet{Sun2025}.
The goal would be to determine which models become too complex and fail to align the curves effectively, and which ones lead to better alignment of the curves, 
in order to understand the critical point of breakdown between alignment and smoothing. 

Further future work can undoubtedly concern the extension of our prior elicitation strategy, particularly regarding landmarks. The idea is to introduce a methodology yielding an innovative hybrid approach: in the current literature -- particularly within the frequentist framework -- landmarks are typically fixed as inputs to the model; in contrast, this hybrid strategy incorporates fixed landmarks as the prior mean within a Bayesian model, with a suitably tuned variance. In this way, the Bayesian model is centered around the classical frequentist landmarks while allowing for data-driven deviations from them. This proposal presents both methodological and computational challenges that are part of our ongoing research.

\section*{Acknowledgments}
We wish to thank the U-motion laboratory at Ume\aa~University and Charlotte K. Hager for providing the knee flexion data, which played an essential role in this research.

\bibliographystyle{apalike}
\bibliography{bibliography}

\newpage

\appendix
\section{Proofs and further details on the proposed model}
\subsection{Proof of Proposition 1}
\label{Beta proof}
We will prove that under the novel prior for the warping parameters the prior distribution of each component of the vector $\boldsymbol{\phi_{gi}}$ follows a beta distribution. The result clearly does not apply to the first (that is equal to 0 surely) and to the last component (that is equal to 1 surely).
Without loss of generality, thanks to the a \emph{priori} independence of the warping vectors, we just need to prove the result for a fixed individual i of a fixed group g. 

It is straightforward to observe that
\begin{align*}
    T_{gi} &= \sum_{k=1}^{q}  \xi_{gik} \sim gamma(\sum_{k=2}^{q} a_k,b),\\ 
    V_{gij} &= \sum_{k=1}^{j} \xi_{gik} \sim gamma(\sum_{k=2}^{j} a_k,b),\\ 
    T_{gi} - V_{gij} & = \sum_{k=j+1}^{q} \xi_{gik} \sim gamma(\sum_{k=j+1}^{q} a_k,b).
\end{align*}
But since $ V_{gij}$ and $ (T_{gi} - V_{gij}) $ are functions of independent random variables it follows that $ V_{gij} \perp (T_{gi} - V_{gij}) $.
So we can rewrite $\phi_{ij}$ as 
\begin{align*}
    \phi_{gij} &= \frac{\sum_{k=1}^{j} \xi_{gik}}{T_gi} = \frac{\sum_{k=1}^{j} \xi_{gik}}{T - V_{ij} + V_{ij}}\\ 
    &= \frac{\sum_{k=1}^{j} \xi_{gik} }{(T_{gi}- V_{gij}) + V_{gij}} = \frac{V_{gij} }{(T_{gi} - V_{gij}) + V_{gij}}. \\
\end{align*}
But this is the definition of a beta random variable. Therefore we get 
\begin{align*}
    \phi_{gij} &\sim Beta(\alpha_{1j},\alpha_{2j}) \qquad j=2,\dots,q-1 \qquad i=1,\dots,N_g, \qquad g=1,...,G,\\
    \alpha_{1j}  &= \sum_{k=2}^{j} a_k \qquad j=2,\dots,q-1,\\
        \alpha_{2j} &= \sum_{k=j+1}^{q} a_k \qquad j=2,\dots,q-1.\\
\end{align*}
Then using the well know results for a beta we can compute the prior mean of $\phi_{gij}$ $\forall j=2,...,q-1$
\begin{align*}
    \mathbb{E}[\phi_{gij}] &= \frac{\alpha_{1j}}{\alpha_{1j} + \alpha_{2j}} = 
    \frac{\sum_{k=2}^{j} a_k}{\sum_{k=2}^{j} a_k + \sum_{k=j+1}^{q} a_k}
    = \frac{\sum_{k=2}^{j} a_k}{\sum_{k=2}^{q} a_k}. \\
\end{align*}
as well as the prior variance
\begin{align*}
    Var(\phi_{gij}) & = \frac{\alpha_{1j}\alpha_{2j}}{(\alpha_{1j}+\alpha_{1j}-1)(\alpha_{1j}+ \alpha_{2j})^2}\\
    &= \frac{(\sum_{k=2}^{j} a_k)(\sum_{k=j+1}^{q} a_k)}{(\sum_{k=2}^{j} a_k+\sum_{k=j+1}^{q} a_k+1)(\sum_{k=2}^{j} a_k+\sum_{k=j+1}^{q} a_k)^2}\\ 
    &= \frac{(\sum_{k=2}^{j} a_k)(\sum_{k=j+1}^{q} a_k)}{(\sum_{k=2}^{q} a_k+1)(\sum_{k=2}^{q} a_k)^2}.
\end{align*}
Then, using the \emph{a priori} independence of all the $\xi_{gij}$, we can extend this result to all individuals and all groups, concluding the proof. 

We now prove that for any vector $\phi_{gi}$, the law of the vector of increments is a Dirichlet distribution and is independent of $T_{gi}$. Once again, leveraging the prior independence of the variables, it is sufficient to show the result for a single individual from a single group. For the sake of clarity, the indices $i$ and $g$ are omitted in the following proof. 
Starting from the definition of the vector $\phi$, we express the variables $\xi_{j}$ in terms of the components of the vector.

\begin{align*}
    \phi_{2} &= \frac{\xi_{2}}{T} \quad \Rightarrow \quad \xi_{2} =   \phi_{2} T,\\
    \phi_{3} &= \frac{\xi_{2} + \xi_{3}}{T} = \frac{\xi_{2}}{T} + \frac{\xi_{3}}{T} = \phi_{2} + \frac{\xi_{3}}{T} \quad \Rightarrow \quad \xi_{3} = ( \phi_{3} - \phi_{2} )  T,\\
    .\\
    .\\
    \phi_{q} &= \frac{\xi_{2} + \cdots + \xi_{q}}{T} = \frac{\xi_{2} + \cdots + \xi_{q-1}}{T} + \frac{\xi_{q}}{T} = \phi_{q-1} + \frac{\xi_{q}}{T} \quad \Rightarrow \quad \xi_{q} = (1 - \phi_{q-1})  T.\\
\end{align*}
Therefore, we have: 
\begin{align*}
    \xi_{j} &= \begin{cases} 
    \phi_{2}  T & \text{if } j = 2, \\
    ( \phi_{j} - \phi_{j-1} )   T & \text{if } j > 2, \\
    (1 - \phi_{j-1})   T & \text{if } j = q.
    \end{cases}
\end{align*}
Computing the partial derivatives of $\xi_{2}, \dots, \xi_{q}$ with respect to $\phi_{2}, \dots, \phi_{q-1}$ and $T$ we get: 
\begin{itemize}
    \item For j = 2 
   \begin{align*}
    \frac{\partial \xi_{2}}{\partial \phi_{k}} =
    \begin{cases}
        0 & \text{if } k \neq 2 \\
        T & \text{if } k = 2
    \end{cases}, \quad
    \frac{\partial \xi_{2}}{\partial T }= \phi_{2}.
\end{align*}
    \item For $3 \leq j \leq q-1$ 
    \begin{align*}
    \frac{\partial \xi_{j}}{\partial \phi_{k}} = 
    \begin{cases}
        0 & \text{if } k \neq j-1, j \\
        T & \text{if } k = j \\
        -T & \text{if } k = j-1 \\
    \end{cases}, \quad \frac{\partial \xi_{j}}{\partial T} = \phi_{j}-\phi_{j-1}.
\end{align*}
\item For j = q
\begin{align*}
    \frac{\partial \xi_{q}}{\partial \phi_{k}} &=
    \begin{cases}
        0 & \text{if } k \neq q-1 \\
        -T & \text{if } k = q-1
    \end{cases}, \quad
    \frac{\partial \xi_{q}}{\partial T} = 1-\phi_{q-1}.
\end{align*}
\end{itemize}
Thus, the Jacobian matrix of the partial derivatives of $\xi_{2}, \dots, \xi_{gi,q}$ with respect to $\phi_{2}, \dots, \phi_{q-1}, T$ is given by:  
\[
J =
\begin{bmatrix}
\frac{\partial \xi_2}{\partial \phi_2} & \frac{\partial \xi_2}{\partial \phi_3} & \cdots & \frac{\partial \xi_2}{\partial \phi_{q-1}} & \frac{\partial \xi_2}{\partial T} \\
\frac{\partial \xi_3}{\partial \phi_2} & \frac{\partial \xi_3}{\partial \phi_3} & \cdots & \frac{\partial \xi_3}{\partial \phi_{q-1}} & \frac{\partial \xi_3}{\partial T} \\
\vdots & \vdots & \ddots & \vdots & \vdots \\
\frac{\partial \xi_q}{\partial \phi_2} & \frac{\partial \xi_q}{\partial \phi_3} & \cdots & \frac{\partial \xi_q}{\partial \phi_{q-1}} & \frac{\partial \xi_q}{\partial T}
\end{bmatrix}
=
\begin{bmatrix}
T & 0 & \cdots & 0 & \phi_2 \\
-T & T & \cdots & 0 & \phi_3 - \phi_2 \\
0 & -T & \cdots & 0 & \phi_4 - \phi_3 \\
\vdots & \vdots & \ddots & \vdots & \vdots \\
0 & 0 & \cdots & T & 1 - \phi_{q-1}
\end{bmatrix} \in \mathbb{R}^{(q-1) \times (q-1)}.
\]
\noindent
We are interested in calculating the determinant of this matrix.
\begin{equation}
\label{eq:det_J}
\det(J) = (-1)^{q-1+1}\phi_{2} \det(J_{1,q-1}) + \cdots + (-1)^{q-1+q-1} (1 - \phi_{q-1}) \det(J_{q-1,q-1}).
\end{equation}
\noindent
where \( J_{i,j} \) represents the submatrix obtained by removing the \( i \)-th row and \( j \)-th column.
\begin{equation}
\label{eq:det_J_submatrix}
\det(J_{i, q-1}) = (T)^{i-1}  (-T)^{q-1-i} \quad \text{for } i = 1, \dots, q-1.
\end{equation}
Substituting \eqref{eq:det_J_submatrix} into \eqref{eq:det_J}, we obtain:
\begin{align*}
\det(J) &= (-1)^{q-1+1}   \phi_{2}   (T)^{1-1} (-T)^{q-1-1} +\cdots \\
&\cdots + (-1)^{q-1+q-1}   (1 - \phi_{q-1})  (T)^{q-1-1}  (-T)^{q-1-(q-1)}\\
&= (-1)^{2q-2}  \phi_2   (T)^{q-2} +\cdots+ (-1)^{2q-2}  (1 - \phi_{q-1})  (T)^{q-2}\\
& = (-1)^{2q-2}  (T)^{q-2}   (\phi_2 + \phi_3 -\phi_2 + \cdots + 1- \phi_{q-1} )\\ 
& = T^{q-2}.
\end{align*}
At this point, using well-known theoretical results, we can write:
\begin{align*}
&f_{\phi_2, \ldots, \phi_{q-1}, T}(\phi_2, \ldots, \phi_{q-1}, T) = 
f_{\xi_2, \ldots, \xi_q}(\phi_2 T, \ldots, (1 - \phi_{q-1}) T)  |\det(J)|\\
& \stackrel{\perp}{=} f_{\xi2}(\phi_{2}T)  f_{\xi_3}((\phi_{3} - \phi_{2})T)  ...  f_{\xi_q}((1 - \phi_{q-1})T) T^{q-2}\\
&= \frac{b^{a_2}}{\Gamma(a_2)}(\phi_{2}T)^{a_2 - 1} e^{-b\phi_{2}T}\mathcal{I}_{(0,+\infty)}(\phi_{2} T)  ...\\
& \times ... \frac{b^{a_q}}{\Gamma(a_q)}((1-\phi_{q-1})T)^{a_q - 1} e^{-b(1-\phi_{q-1})T}\mathcal{I}_{(0,+\infty)}((1-\phi_{q-1}) T)  T^{q-2}\\
&= \frac{b^{\sum_{j=2}^{q} a_j}}{\prod_{j=2}^{q}\Gamma(a_j)}\phi_2^{a2-1} (\phi_3 - \phi_2 ) ^{a_3-1}  ... (1-\phi_{q-1})^{a_q -1} (T)^{a_2-1+a_3-1 + \ldots + a_q-1+q-2}\\
&\times e^{-bT(\phi_2 + \phi_3 - \phi_2 + \ldots + 1-\phi_{q-1})} \mathcal{I}_{(0,+\infty)}(\phi_{2} T)  ...  \mathcal{I}_{(0,+\infty)}((1-\phi_{q-1}) T)\\
&= \frac{\Gamma(\sum_{j=2}^{q} a_j)}{\Gamma(\sum_{j=2}^{q} a_j)} 
\frac{b^{\sum_{j=2}^{q} a_j}}{\prod_{j=2}^{q}\Gamma(a_j)}\phi_2^{a2-1} (\phi_3 - \phi_2 ) ^{a_3-1}  ...  (1-\phi_{q-1})^{a_q -1}  (T)^{\sum_{j=2}^{q} a_j -(q-1) +q-2} \\
&\times e^{-bT} \mathcal{I}_{(0,+\infty)}(\phi_{2} T)  ... \mathcal{I}_{(0,+\infty)}((1-\phi_{q-1}) T)\\
&= \frac{\Gamma(\sum_{j=2}^{q} a_j)}{\prod_{j=2}^{q}\Gamma(a_j)} \phi_2^{a2-1} (\phi_3 - \phi_2 ) ^{a_3-1}  ...  (1-\phi_{q-1})^{a_q -1}\\
& \times \frac{b^{\sum_{j=2}^{q} a_j}}{\Gamma(\sum_{j=2}^{q} a_j)} (T)^{\sum_{j=2}^{q} a_j -1}  e^{-bT} \mathcal{I}_{(0,+\infty)}(\phi_{2} T)  ...  \mathcal{I}_{(0,+\infty)}((1-\phi_{q-1}) T).
\end{align*}
where $\mathcal{I}_{(0,+\infty)}(t)$ denotes the indicator function of the set $(0,+\infty)$ evaluated at $t$. 
Now, focusing only on the indicator functions and recalling that 
$T$ is defined as the sum of non-negative quantities, we can write:
\begin{align*}
    &\mathcal{I}_{(0,+\infty)}(\phi_2 T) \Rightarrow \phi_2 T \geq 0 \Rightarrow \phi_2 \geq 0 , T \geq 0 \Rightarrow   \mathcal{I}_{(0,+\infty)}(\phi_2) \mathcal{I}_{(0,+\infty)}(T),\\
     &\mathcal{I}_{(0,+\infty)}((\phi_3-\phi_2) T) \Rightarrow \phi_3-\phi_2 \geq 0 , T \geq 0 \Rightarrow   \mathcal{I}_{(0,+\infty)}(\phi_3 -\phi_2)  \mathcal{I}_{(0,+\infty)}(T),\\
     &..,\\
     &\mathcal{I}_{(0,+\infty)}((1-\phi_{q-1}) T) \Rightarrow 1-\phi_{q-1} \geq 0 , T \geq 0 \Rightarrow   \mathcal{I}_{(0,+\infty)}(1 -\phi_{q-1})  \mathcal{I}_{(0,+\infty)}(T).\\
\end{align*}
From this, we obtain that all the increments $\phi_2$,...,$1-\phi_{q-1}$ must be non-negative. Moreover, we know, by construction that they sum to $1$.
Therefore, the product of the indicator functions can be rewritten as:
\begin{equation*}
    \mathcal{I}_{(0,+\infty)}(\phi_{2} T)  ...  \mathcal{I}_{(0,+\infty)}((1-\phi_{q-1}) T) = \mathcal{I}_{S^{q-2}}(\phi_2,\phi_3 - \phi_2,\ldots,1-\phi_{q-1})  \mathcal{I}_{(0,+\infty)}( T).
\end{equation*}
where $S^{q-2}$ is the set given by: 
\begin{equation*}
    S^{q-2} = \left\{ x = (x_1,...,x_{q-1}) \in \mathbb{R}^{q-1} : x_j \geq 0, \sum_{j=1}^{q-1} x_j = 1 \right\}.
\end{equation*}
Using this result, we can rewrite the joint density of $\phi_2,...,\phi_{q-1},T$ as: 
\begin{align*}
&f_{\phi_2, \ldots, \phi_{q-1}, T}(\phi_2, \ldots, \phi_{q-1}, T) = 
     \frac{\Gamma(\sum_{j=2}^{q} a_j)}{\prod_{j=2}^{q}\Gamma(a_j)} \phi_2^{a2-1} (\phi_3 - \phi_2 ) ^{a_3-1}  ...  (1-\phi_{q-1})^{a_q -1}\\
& \times \mathcal{I}_{S^{q-2}}(\phi_2,\phi_3 - \phi_2,\ldots,1-\phi_{q-1})   \frac{b^{\sum_{j=2}^{q} a_j}}{\Gamma(\sum_{j=2}^{q} a_j)}  (T)^{\sum_{j=2}^{q} a_j -1}  e^{-bT} \mathcal{I}_{(0,+\infty)}(T)\\
&= f_{Dir(a_2,...,a_q)}(\phi_2,\ldots,1-\phi_{q-1})  f_{Gamma(\sum_{j=2}^{q} a_j,b)}(T).
\end{align*}
where $f_{Dir(a_2,...,a_q)}$ and $f_{Gamma(\sum_{j=2}^{q} a_j,b)}$ denote the densities of a Dirichlet distribution with parameters $a_2,...,a_q$ and of a Gamma distribution of parameters $\sum_{j=2}^{q} a_j,b$, respectively. 
From this result it is clear that:
\begin{itemize}
    \item $(\phi_2,\phi_3-\phi2,...,1-\phi_{q-1}) \sim Dir(a_2,...,a_q)$.
    \item $T \sim gamma(\sum_{j=2}^{q} a_j,b)$.
    \item $T \perp \phi_j \quad j=2,...,q-1$.
\end{itemize}
which conclude the proof. 

\subsection{Computation of the covariance matrix for the warping vectors}\label{app:covariance}
We are now interested in calculating the prior covariance matrix for the warping vectors $\phi_{gi}$. Once again, leveraging the prior independence of the variables, it is sufficient to show the result for a single individual from a single group. For the sake of clarity the indices $i$ and $g$ are omitted in the following proof.
Since \( \phi_1 = 0 \), it is trivial to observe that \( \text{Cov}(\phi_1, \phi_j) = 0 \) for every \( j = 1, \dots, q \). Similarly, since \( \phi_q = 1 \), we obtain  \( \text{Cov}(\phi_q, \phi_j) = 0 \) for every \( j = 1, \dots, q \).
Additionally, using the result i of proposition \ref{prop:marginal} regarding the marginal law of the components, it is easy to observe that:
\[
\text{Cov}(\phi_j, \phi_j) = \text{Var}(\phi_j) =
\begin{cases} 
0 & \text{if } j = 1,q, \\
\frac{\sum_{k=2}^{j} a_k  \sum_{k=j+1}^{q} a_k}{(\sum_{k=2}^{q} a_k +1)(\sum_{k=2}^{q} a_k)^2}
 & \text{if } 2 \leq j \leq q-1.
\end{cases}
\]
To calculate the covariance between \(\phi_i\) and \(\phi_j\), we instead need to use a preliminary result. Let use introduce a new set of variables $X_2,...,X_q$ defined as: 
\[
X_j =
\begin{cases}
\phi_2 & \text{if } j = 2, \\
\phi_j - \phi_{j-1} & \text{if } 3 \leq j \leq q-1, \\
1 - \phi_{q-1} & \text{if } j = q.
\end{cases}
\]
The Jacobian matrix of the partial derivatives of \(\phi_2, \phi_3 - \phi_2, \dots, 1 - \phi_{q-1}\) with respect to \(x_2, \dots, x_q\) is exactly the identity matrix. Therefore it is trivial to observe that: 
\begin{equation*}
    f_{X_2,\ldots,X_q}(X_2,...,X_q)  = f_{\phi_2,\phi_3-\phi_2,...,1-\phi_{q-1}}(X_2,...,X_q) = f_{Dir(a_2,...,a_q)}(X_2,...,X_q).
\end{equation*}
and so $(X_2,...,X_q) \sim Dir(a_2,...,a_q)$.
Moreover we can observe that each each component $\phi_j$ can be written as a function of \(X_2, \dots, X_q\):
\begin{align*}
    \phi_{2} &= X_2,\\
    X_{3} &= \phi_3 -\phi_2 \quad \Rightarrow \quad \phi_{3} =  X_{3} + \phi_2 = X_{3} + X_{2},\\
    ...,\\
    \phi_{q-1} &= 1 - X_q = \sum_{k=2}^{q-1} X_{k}.\\
\end{align*}
which is: 
\begin{equation*}
    \phi_j = \sum_{k=2}^{j} X_k.
\end{equation*}
Using these two results, we can compute \(\text{Cov}(\phi_i, \phi_j)\) for any \(i, j \in \{2, \dots, q-1\}\).
We consider only the case \(i < j\); the other one follows naturally by symmetry. 
\begin{align*}
    &\text{Cov}( \phi_i, \phi_j ) = \text{Cov}( \sum_{l=2}^{i} X_l, \sum_{k=2}^{j} X_k ) =  \sum_{l=2}^{i} \sum_{k=2}^{j} \text{Cov}(X_l,X_k)\\
    &= \sum_{l=2}^{i}( \text{Cov}(X_l,X_2) + \text{Cov}(X_l,X_3) + \ldots + \text{Cov}(X_l,X_j) )\\
    &= \sum_{l=2}^{i}( \text{Cov}(X_l,X_l) + \sum_{\substack{k=2 \\ k \neq l}}^j \text{Cov}(X_l,X_k) )\\
    & = \sum_{l=2}^{i}( \frac{ a_l  ( \sum_{m=2}^{q}a_m - a_l)}{(\sum_{m=2}^{q}a_m +1)  (\sum_{m=2}^{q}a_m)^2} -  \sum_{\substack{k=2 \\ k \neq l}}^j \frac{a_l  a_k}{(\sum_{m=2}^{q}a_m +1)  (\sum_{m=2}^{q}a_m)^2} )\\
    &= \sum_{l=2}^{i} \frac{a_l}{(\sum_{m=2}^{q}a_m +1) (\sum_{m=2}^{q}a_m)^2}  ( \sum_{m=2}^{q}a_m - a_l -\sum_{\substack{k=2 \\ k \neq l}}^j a_k )\\
    &=  \sum_{l=2}^{i} \frac{a_l}{(\sum_{m=2}^{q}a_m +1)}  ( \sum_{k=j+1}^q a_k) = \frac{\sum_{l=2}^i a_l  \sum_{k=j+1}^q a_k}{(\sum_{m=2}^{q}a_m +1)  (\sum_{m=2}^{q}a_m)^2}.
\end{align*}



\subsection{Computation of the Full Conditionals}
\label{Full cond count}
In this section, we report the computations of the full conditional distributions. For the sake of clarity of notation, we once again assume that all curves have the same number of observations. However, the generalization to cases with different numbers of observations is both possible and straightforward.  

\subsubsection{Likelihood}
Before presenting the calculations of the full conditional distributions for the parameters, we introduce the likelihood of the data given all the parameters, which forms the basis of these computations, making them clearer.

The likelihood expression is given by:

\begin{align*}
     f(Y \mid B,\Gamma,\Phi,\sigma_\epsilon^2) &= \prod_{g=1}^{G}\prod_{i=1}^{N_g} f(Y_{gi} \mid B,\Gamma,\Phi,\sigma_\epsilon^2)\\
    & = \prod_{g=1}^{G}\prod_{i=1}^{N_g} \mathcal{N}_{n_{gi}}(\tilde{m}_{gi},\sigma_\epsilon^2\mathbb{I}_{n_{gi}}) \propto \prod_{g=1}^{G}\prod_{i=1}^{N_g} \frac{1}{(\sigma_\epsilon^2)^{\frac{n_{gi}}{2}}}e^{-\frac{1}{2\sigma_\epsilon^2}(Y_{gi}-\tilde{m}_{gi})^T(Y_{gi}-\tilde{m}_{gi})}\\
    & = \frac{1}{(\sigma_\epsilon^2)^{\sum_{g=1}^{G}\sum_{i=1}^{N_g}\frac{n_{gi}}{2}}}e^{-\frac{1}{2\sigma_\epsilon^2}\sum_{g=1}^{G}\sum_{i=1}^{N_g}(Y_{gi}-\tilde{m}_{gi})^T(Y_{gi}-\tilde{m}_{gi})}.
\end{align*}

\subsubsection{Full Conditional of \protect{$\beta_g$}}

Due to the a \emph{priori} independence of $\boldsymbol{\beta_g}$ for g=1,...,G we have that conditionally to all the other parameters they are a \emph{posteriori} independent. 
The computations of the full conditional for each $\boldsymbol{\beta_g}$ for $g = 1, \ldots, G$ easily follow from
those for the full conditional distribution of the unique common shape parameter $\beta$ in the model by Telesca and Inoue (2008) \cite{Telesca}. 
The only differences are that the term $c_i 1_{n_i} \in \mathbb{R}^{n_i}$ 
is now replaced by the term $\mathcal{B}_{\gamma} \boldsymbol{\gamma_{gi}} \in \mathbb{R}^{n_{gi}}$, all the $a_i$ coefficients are now set to equal to 1 and that the matrix $X$ and the vector $Y$ now account only for the elements of the specific group $g$.

Therefore we easily get that the posterior distribution of $\boldsymbol{\beta_g}$ for g=1,...,G is
\vspace{12pt}
\begin{align*}
\pi(\boldsymbol{\beta_g}| -) & \sim \mathcal{N}_p (\boldsymbol{m_{\beta_g}}, \boldsymbol{V_{\beta_g}}),\\
\boldsymbol{V^{-1}_{\beta_g} } & = [ \boldsymbol{\Sigma^{-1}_{\beta}} + \frac{1}{\sigma_{\epsilon}^2} \boldsymbol{X_g^{T}}\boldsymbol{X_g}],\\
\boldsymbol{m_{\beta_g} } & = \boldsymbol{V_{\beta_g} }[ \frac{1}{\sigma_{\epsilon}^2} \boldsymbol{X_g^{T}}(\boldsymbol{Y_g} -\boldsymbol{C_{\gamma_g}})],\\ 
\boldsymbol{C_{\gamma_g}} & = (\mathcal{B}_{\beta}^T(h_1(t_{g1},\phi_{g1}))\gamma_{g1},...,\mathcal{B}_{\beta}^T(h_{N_g}(t_{gN_g},\phi_{N_g}))\gamma_{N_g}),\\
\boldsymbol{X_g} & = (\mathcal{B}_{\beta}^T(h_{g1}(t_{g1},\phi_{g1})),...,\mathcal{B}_{\beta}^T(h_{gN_g}(t_{gN_g},\phi_{gN_g}))).\\
\end{align*}
 
\subsubsection{Full Conditional of \protect{$\lambda$}}

\begin{align*}
\lambda \mid \text{rest} &\propto f(B \mid \lambda) f(\lambda \mid a_\lambda, b_\lambda) \propto \prod_{g=1}^{G} N_p(0, \Sigma_{\beta}) f(\lambda \mid a_\lambda, b_\lambda)\\
    &\propto \prod_{g=1}^{G} \frac{1}{\lambda^{\frac{p}{2}}} e^{-\frac{1}{2\lambda} \boldsymbol{\beta_g}^T \Omega \boldsymbol{\beta_g}} \frac{e^{-\frac{b_\lambda}{\lambda}}}{\lambda^{a_\lambda+1}} = \frac{1}{\lambda^{\frac{Gp}{2}+a\lambda+1}}e^{-\frac{1}{\lambda}(b_\lambda + \frac{1}{2} \sum_{g=1}^{G} \boldsymbol{\beta_g}^T \Omega \boldsymbol{\beta_g}}).
\end{align*}
But it is the kernel of a inv-gamma distribution. 
Therefore we have 
\begin{align*}
\pi(\lambda| -) & \sim  inv-gamma (a_{\lambda}^*,b_{\lambda}^*),\\
a_{\lambda}^{*} &= a_{\lambda} +  \frac{Gp}{2},\\
b_{\lambda}^{*}& = b_{\lambda} +  \frac{1}{2}\sum_{g=1}^{G} \boldsymbol{\beta_g}^T \Omega \boldsymbol{\beta_g}.
\\
\end{align*}

\subsubsection{Full Conditional of \protect{$\gamma_{gi}$}}
Once again using the a \emph{priori} independence of $\boldsymbol{\gamma_{gi}}$ for  $i=1,...,N_g$, g=1,...,G we have that conditionally to all the other parameters they are a \emph{posteriori} independent. 
Moreover also in this case the computations of the full conditional for each $\boldsymbol{\gamma_{gi}}$ for  $i=1,...,N_g$, g=1,...,G easily follow from
those for the full conditional distribution of the unique common shape parameter $\beta$ in the model by Telesca and Inoue (2008) \cite{Telesca}. 
The only differences are that the term $c_i 1_{n_i} \in \mathbb{R}^{n_i}$ 
is now replaced by the term $\mathcal{B}_{\beta} \beta_{g} \in \mathbb{R}^{n_{gi}}$,all the $a_i$ coefficients are now set to equal to 1, $X = \mathcal{B}_\gamma^T \boldsymbol{\gamma_{gi}}$ and $Y = Y_{gi}$.

Therefore we easily get that the posterior distribution of $\boldsymbol{\gamma_{gi}}$ for  $i=1,...,N_g$, g=1,...,G is 
\vspace{12pt}
\begin{align*}
\pi(\boldsymbol{\gamma_{gi}}| -) &\overset{\text{ind}}{\sim} \mathcal{N}_k (\boldsymbol{m_{\gamma_{gi}}}, \boldsymbol{V_{\gamma_{gi}}}),\\
    \boldsymbol{V^{-1}_{\gamma_{gi}} } & = \left[ \boldsymbol{\Sigma^{-1}_{\gamma}} + \frac{1}{\sigma_{\epsilon}^2} \mathcal{B}_{\gamma}(h_{gi}(\boldsymbol{t_{gi}},\boldsymbol{\phi_{gi}})) \mathcal{B}_{\gamma}^T(h_{gi}(\boldsymbol{t_{gi}},\boldsymbol{\phi_{gi}})) \right],\\
    \boldsymbol{m_{\gamma_{gi}} } & = \boldsymbol{V_{\gamma_{gi}} } \left[ \frac{1}{\sigma_{\epsilon}^2} \mathcal{B}_{\beta}(h_{gi}(\boldsymbol{t_{gi}},\boldsymbol{\phi_{gi}}))(\boldsymbol{Y_{gi}} - \boldsymbol{C_{\beta_g}^{gi}}) \right],\\
    \boldsymbol{C_{\beta_g}^{gi}} & = \mathcal{B}_{\beta}^T(h_{gi}(\boldsymbol{t_{gi}},\boldsymbol{\phi_{gi}}))\boldsymbol{\beta_g}.
\end{align*}

\subsubsection{Full Conditional of \protect{$\sigma_{\gamma}^2$}}
\begin{align*}
    \sigma_{\gamma}^2 \mid rest & \propto f(\Gamma \mid \sigma_{\gamma}^2) f(\sigma_{\gamma}^2 \mid a_\gamma,b\gamma) = \prod_{g=1}^{G}\prod_{i=1}^{N_g} f(\boldsymbol{\gamma_{gi}} \mid \sigma_{\gamma}^2) f(\sigma_{\gamma}^2 \mid a_\gamma,b\gamma)\\ 
   & \propto \prod_{g=1}^{G}\prod_{i=1}^{N_g} \frac{1}{(\sigma_{\gamma}^2)^\frac{k}{2}}e^{-\frac{1}{2\sigma_{\gamma}^2}\boldsymbol{\gamma_{gi}}^T\mathbb{I}_k\boldsymbol{\gamma_{gi}}}\frac{e^{-\frac{b_\gamma}{\sigma_{\gamma}^2}}}{(\sigma_{\gamma}^2)^{a_\gamma+1}}\\
   & = \frac{e^{-\frac{1}{2\sigma_\gamma^2}\sum_{g=1}^{G}\sum_{i=1}^{N_g}\boldsymbol{\gamma_{gi}}^T\boldsymbol{\gamma_{gi}}}}{(\sigma_{\gamma}^2)^{\sum_{g=1}^{G}\sum_{i=1}^{N_g}\frac{k}{2}}}\frac{e^{-\frac{b_\gamma}{\sigma_{\gamma}^2}}}{(\sigma_{\gamma}^2)^{a_\gamma+1}}\\
   & = \frac{e^{-\frac{1}{2\sigma_\gamma^2}\sum_{g=1}^{G}\sum_{i=1}^{N_g}\boldsymbol{\gamma_{gi}}^T\boldsymbol{\gamma_{gi}}}}{(\sigma_{\gamma}^2)^{\frac{kN}{2}}}\frac{e^{-\frac{b_\gamma}{\sigma_{\gamma}^2}}}{(\sigma_{\gamma}^2)^{a_\gamma+1}}\\
   & = \frac{e^{-\frac{1}{\sigma_{\gamma}^2}(b_\gamma + \frac{1}{2}\sum_{g=1}^{G}\sum_{i=1}^{N_g}\boldsymbol{\gamma_{gi}}^T\boldsymbol{\gamma_{gi}} )}} {(\sigma_{\gamma}^2)^{\frac{kN}{2} + a_\gamma +1}}.
\end{align*}
But it is the kernel of a inv-gamma distribution. 
Therefore we have 
\begin{align*}
\pi(\sigma_\gamma^2| -) & \sim  inv-gamma (a_{\gamma}^*,b_{\gamma}^*),\\
a_{\gamma}^{*} &= a_{\gamma} +  \frac{Nk}{2},\\
b_{\gamma}^{*}& = b_{\gamma} + \frac{1}{2} \sum_{g=1}^{G}\sum_{i=1}^{N_g}\boldsymbol{\gamma_{gi}}^T\boldsymbol{\gamma_{gi}}.
\\
\end{align*}

\subsubsection{Full conditional of \protect{$\sigma_{\epsilon}^2$}}
\begin{align*}
    \sigma_{\epsilon}^2 \mid rest & \propto f(Y \mid B,\Gamma,\Phi,\sigma_{\epsilon}^2) f(\sigma_{\epsilon}^2 \mid a_\epsilon,b_\epsilon)\\
    &= \prod_{g=1}^{G} \prod_{i=1}^{N_g} f(Y_{gi} \mid \boldsymbol{\beta_g}, \boldsymbol{\gamma_{gi}},\boldsymbol{\phi_{gi}},\sigma_{\epsilon}^2) f(\sigma_{\epsilon}^2 \mid a_\epsilon,b_\epsilon)\\ 
    &= \prod_{g=1}^{G} \prod_{i=1}^{N_g} N_{n_{gi}}(\tilde{m}_{gi},\sigma_{\epsilon}^2I_{n_{gi}}) f(\sigma_{\epsilon}^2 \mid a_\epsilon,b_\epsilon)\\
    & \propto \prod_{g=1}^{G} \prod_{i=1}^{N_g} \frac{1}{(\sigma_{\epsilon}^2)^\frac{n_{gi}}{2}} e^{-\frac{1}{2\sigma_{\epsilon}^2}(Y_{gi}-\tilde{m}_{gi})^T(Y_{gi}-\tilde{m}_{gi})}f(\sigma_{\epsilon}^2 \mid a_\epsilon,b_\epsilon)\\
    &\propto \frac{e^{-\frac{1}{2\sigma_{\epsilon}^2}\sum_{g=1}^{G} \sum_{i=1}^{N_g}(Y_{gi}-\tilde{m}_{gi})^T(Y_{gi}-\tilde{m}_{gi})}}{(\sigma_\epsilon^2)^{\frac{1}{2} \sum_{g=1}^{G} \sum_{i=1}^{N_g}n_{gi}}} \frac{e^{-\frac{b_\epsilon}{\sigma_\epsilon^2}}}{(\sigma_\epsilon^2)^{a_\epsilon + 1}}\\
    &= \frac{e^{-\frac{1}{\sigma_{\epsilon}^2}(b_\epsilon+ \frac{1}{2}\sum_{g=1}^{G} \sum_{i=1}^{N_g}(Y_{gi}-\tilde{m}_{gi})^T(Y_{gi}-\tilde{m}_{gi})}}{(\sigma_\epsilon^2)^{a_\epsilon+ \frac{1}{2} \sum_{g=1}^{G} \sum_{i=1}^{N_g}n_{gi}+1}} .
\end{align*}
But this is the kernel of a inv-gamma distribution and therefore we get 
\begin{align*}
    \pi(\sigma^2_{\epsilon}| -) &\sim inv-gamma(a_{\epsilon}^*,b_{\epsilon}^*),\\
    a_{\epsilon}^*&=a_{\epsilon} +  \frac{1}{2}\sum_{g=1}^{G} \sum_{i=1}^{N_g} n_{gi},\\
    b_{\epsilon}^*&= b_{\epsilon} +  \frac{1}{2}\sum_{g=1}^{G} \sum_{i=1}^{N_g} (\boldsymbol{Y}_{gi} - \boldsymbol{\tilde{m}}_{gi})^{T} (\boldsymbol{Y}_{gi} - \boldsymbol{\tilde{m}}_{gi}).\\
   \boldsymbol{\tilde{m}}_{gi}&=\mathcal{B}_{\beta}(h_{gi}(\boldsymbol{t_{gi}},\boldsymbol{\phi_{gi}}))^T \boldsymbol{\beta_g} + \mathcal{B}_{\gamma}(h_{gi}(\boldsymbol{t_{gi}},\boldsymbol{\phi_{gi}}))^T \boldsymbol{\gamma}_{gi}     \qquad   i=1,...,N_g, \qquad g=1,...,G.\\
\end{align*}

\subsubsection{Full conditional of the warping parameters}
\label{sec:full codntional csi}
The full conditional distribution of $\boldsymbol{\phi_{gi}}$ is expressed as a function of $\boldsymbol{\xi_{gi}}$, as $\boldsymbol{\phi_{gi}}$ can be deterministically computed once $\xi$ is known.
\begin{align*}
    \boldsymbol{\xi_{gi}} \mid rest & \propto f(Y \mid B,\Gamma,\Phi,\sigma_{\epsilon}^2) f(\boldsymbol{\xi_{gi}} \mid a_1,...,a_q,b)\\
    &= \prod_{g=1}^{G} \prod_{i=1}^{N_g} f(Y_{gi} \mid \boldsymbol{\beta_g}, \boldsymbol{\gamma_{gi}},\boldsymbol{\xi_{gi}},\sigma_{\epsilon}^2) f(\boldsymbol{\xi_{gi}} \mid a_1,...,a_q,b)\\ 
    &= \prod_{g=1}^{G} \prod_{i=1}^{N_g} N_{n_{gi}}(\tilde{m}_{gi},\sigma_{\epsilon}^2I_{n_{gi}}) f(\boldsymbol{\xi_{gi}} \mid a_1,...,a_q,b)\\
    & \propto \prod_{g=1}^{G} \prod_{i=1}^{N_g}  e^{-\frac{1}{2\sigma_{\epsilon}^2}(Y_{gi}-\tilde{m}_{gi})^T(Y_{gi}-\tilde{m}_{gi})}f(\boldsymbol{\xi_{gi}} \mid a_1,...,a_q,b)\\
    &\propto e^{-\frac{1}{2\sigma_{\epsilon}^2}\sum_{g=1}^{G} \sum_{i=1}^{N_g}(Y_{gi}-\tilde{m}_{gi})^T(Y_{gi}-\tilde{m}_{gi})}  f(\boldsymbol{\xi_{gi}} \mid a_1,...,a_q,b) \\
    &= e^{-\frac{1}{2\sigma_{\epsilon}^2}\sum_{g=1}^{G} \sum_{i=1}^{N_g}(Y_{gi}-\tilde{m}_{gi})^T(Y_{gi}-\tilde{m}_{gi})} \prod_{j=1}^{2} \xi_{g_{ij}}^{a_j - 1} \, e^{-b \xi_{g_{ij}}}.
\end{align*}
where 
\begin{align*}
       \tilde{m}_{gi} &= \mathcal{B}^T_{\beta}( \mathcal{B}_{h}^T(t_{n_{gi}}) \boldsymbol{\phi_{gi}} ) \beta_g + \mathcal{B}^T_{\gamma}( \mathcal{B}_{h}^T(t_{n_{gi}}) \boldsymbol{\phi_{gi}} ) \gamma_{gi},\\
       \boldsymbol{\phi_{gi}} & = (0, \frac{\xi_{gi2} }{\sum_{j=2}^{q} \xi_{gij}},....,\frac{ \sum_{j=2}^{q-1} \xi_{gij} }{\sum_{j=2}^{q} \xi_{gij}}, 1).
\end{align*}
Since is not possible to sample directly from this full conditional distribution, so we used a Metropolis-Hastings step.

\subsection{Further details on the MH step}
\label{MH appendix}
In this section are reported more details and information about the MH step for the warping parameters of the MCMC. 
\subsubsection{The computation of the acceptance probability}
\label{Acceptance probability}
The acceptance probability, $\alpha(\xi_{gij}^{old}, \xi_{gij}^{new})$, of a candidate of point $\xi_{gij}^{new}$ of our Metropolis-Hastings step is the usual one, whose expression is given by: 
{\large
\begin{align*}
    \alpha(\xi_{gij}^{old}, \xi_{gij}^{new}) &= \min\left(1,\frac{\pi(\xi_{gij}^{new} \mid \text{rest})\pi_{prop}(\xi_{gij}^{old})}{\pi(\xi_{gij}^{old} \mid \text{rest})\pi_{prop}(\xi_{gij}^{new})}\right)\\
    & = \min\left(1,\frac{f(Y_{gi} \mid \xi_{gij}^{new},\boldsymbol{\beta_g},\boldsymbol{\gamma_{gi}},\sigma_\epsilon^2)f(\xi_{gij}^{new} \mid a_j,b)\pi_{prop}(\xi_{gij}^{old})}{f(Y_{gi} \mid \xi_{gij}^{old},\boldsymbol{\beta_g},\boldsymbol{\gamma_{gi}},\sigma_\epsilon^2)f(\xi_{gij}^{old} \mid a_j,b)\pi_{prop}(\xi_{gij}^{new})}\right)\\
    & = \min\left(1, \frac{
    \mathcal{N}_{n_{gi}}(\tilde{m}_{gi}(\xi_{gij}^{new}),\sigma_\epsilon^2\mathbb{I}_{n_{gi}})f(\xi_{gij}^{new} \mid a_j,b)\pi_{prop}(\xi_{gij}^{old})}{\mathcal{N}_{n_{gi}}(\tilde{m}_{gi}(\xi_{gij}^{old}),\sigma_\epsilon^2\mathbb{I}_{n_{gi}})f(\xi_{gij}^{old} \mid a_j,b)\pi_{prop}(\xi_{gij}^{new})}\right)\\
    &= \min\left(1,\frac{e^{-\frac{1}{2\sigma\epsilon^2}(Y_{gi}-\tilde{m}_{gi}(\xi_{gij}^{new}))^T(Y_{gi}-\tilde{m}_{gi}(\xi_{gij}^{new}))}  \frac{e^{-\frac{b}{\xi_{gij}^{new}}}}{(\xi_{gij}^{new})^{a_j+1}}
    \frac{e^{-\frac{b_{prop}}{\xi_{gij}^{old}}}}{(\xi_{gij}^{old})^{a_{prop}+1}}} {e^{-\frac{1}{2\sigma\epsilon^2}(Y_{gi}-\tilde{m}_{gi}(\xi_{gij}^{old}))^T(Y_{gi}-\tilde{m}_{gi}(\xi_{gij}^{old}))}  \frac{e^{-\frac{b}{\xi_{gij}^{old}}}}{(\xi_{gij}^{old})^{a_j+1}}
    \frac{e^{-\frac{b_{prop}}{\xi_{gij}^{new}}}}{(\xi_{gij}^{new})^{a_{prop}+1}}}
    \right).
\end{align*}
}
where $\tilde{m}_{gi}(\xi_{gij}^{...})$ is used to emphasize the dependence of $\tilde{m}_{gi}$ on $\xi_{gij}^{...}$. 

In order to avoid numerical instability problems, instead of calculating $\alpha(\xi_{gij}^{old}, \xi_{gij}^{new})$ using the densities, we use the log densities. This approach helps to maintain numerical stability and prevent issues that can arise from working with very small probability values.

\subsection{Model in the case of absence of groups}
\label{No groups}
As previously mentioned, the model is particularly advantageous when the data are grouped. Indeed the smoothing part includes a specific term that considers group dependencies, enabling efficient processing of all data simultaneously and reducing computational load. However, the model can still be used even in the absence of groups or their information.
In that case we simply drop the dependency from the groups obtaining the following smoothing model : 
\begin{equation*}
     m_{i}(t) = m_{i}(t,\boldsymbol{\beta},\boldsymbol{\gamma}_{i}) =  \mathcal{B}_{\beta}(t)^T \boldsymbol{\beta} + \mathcal{B}_{\gamma}(t)^T \boldsymbol{\gamma}_{i} \quad i=1,\ldots,N  \quad t \in [t_0,t_f].
\end{equation*}
The curves are now modeled as the sum of two curves, one that remains individual specific and the other that is shared between all the individuals. 
The warping part remains unchanged, as all terms are either specific to individuals or shared among all individuals already in the formulation that incorporates groups.

\subsubsection{Prior in absence of groups}
The prior distribution remains almost unchanged in the case of no grouped data. 
Indeed the only part that changes is the one related to the common term $\beta$, that now becomes: 
\begin{align*}
    \boldsymbol{\beta} | \lambda &\sim \mathcal{N}_p (\boldsymbol{0}, \boldsymbol{\Sigma_{\beta}}),\\
    \lambda &\sim inv-gamma(a_{\lambda}, b_{\lambda}).
\end{align*}

\subsubsection{Full conditionals in absence of groups}
The computations for the full conditional are identical to those for grouped data, which are reported in Appendix A section \ref{Full cond count}. Thus, we only report the final result. For the full codnitionals in closed form we have: 
\vspace{12pt}
\begin{itemize}
\item For the group-specific shape parameter  $\boldsymbol{\beta}$:
\begin{align*}
\pi(\boldsymbol{\beta}| -) & \sim \mathcal{N}_p (\boldsymbol{m_{\beta}}, \boldsymbol{V_{\beta}}),\\
\boldsymbol{V^{-1}_{\beta} } & = [ \boldsymbol{\Sigma^{-1}_{\beta}} + \frac{1}{\sigma_{\epsilon}^2} \boldsymbol{X^{T}}\boldsymbol{X}], \\
\boldsymbol{m_{\beta} } & = \boldsymbol{V_{\beta} }[ \frac{1}{\sigma_{\epsilon}^2} \boldsymbol{X^{T}}(\boldsymbol{Y} -\boldsymbol{C_{\gamma}})] ,\\ 
\boldsymbol{C_{\gamma}} & = (\mathcal{B}_{\beta}^T(h_1(t_{1},\phi_{g1}))\gamma_{1},...,\mathcal{B}_{\beta}^T(h_{N}(t_{N},\phi_{N}))\gamma_{N}) ,\\
\boldsymbol{X} & = (\mathcal{B}_{\beta}^T(h_{1}(t_{1},\phi_{1})),...,\mathcal{B}_{\beta}^T(h_{N}(t_{N},\phi_{N}))).\\
\end{align*}
\item For the common shape variance parameter $\boldsymbol{\lambda}$:
\begin{align*}
\pi(\lambda| -) & \sim  inv-gamma (a_{\lambda}^*,b_{\lambda}^*),\\
a_{\lambda}^{*} &= a_{\lambda} +  \frac{p}{2},\\
b_{\lambda}^{*}& = b_{\lambda} +  \frac{1}{2} \beta^T \Omega \beta.
\end{align*}
\item
For the individual-specific shape parameters $\gamma_{i}$:
\begin{align*}
\pi(\boldsymbol{\gamma_{i}}| -) &\overset{\text{ind}}{\sim} \mathcal{N}_k (\boldsymbol{m_{\gamma_{i}}}, \boldsymbol{V_{\gamma_{i}}}) \qquad &i=1,...,N,\\
    \boldsymbol{V^{-1}_{\gamma_{i}} } & = \left[ \boldsymbol{\Sigma^{-1}_{\gamma}} + \frac{1}{\sigma_{\epsilon}^2} \mathcal{B}_{\gamma}(h_{i}(\boldsymbol{t_{i}},\boldsymbol{\phi_{i}})) \mathcal{B}_{\gamma}^T(h_{i}(\boldsymbol{t_{i}},\boldsymbol{\phi_{i}})) \right] \qquad &i=1,...,N,\\
    \boldsymbol{m_{\gamma_{i}} } & = \boldsymbol{V_{\gamma_{i}} } \left[ \frac{1}{\sigma_{\epsilon}^2} \mathcal{B}_{\beta}(h_{i}(\boldsymbol{t_{i}},\boldsymbol{\phi_{i}}))(\boldsymbol{Y_{i}} - \boldsymbol{C_{\beta}^{i}}) \right] \qquad &i=1,...,N,\\
    \boldsymbol{C_{\beta}^{i}} & = \mathcal{B}_{\beta}^T(h_{i}(\boldsymbol{t_{i}},\boldsymbol{\phi_{i}}))\beta \qquad &i=1,...,N.
\end{align*}
\item 
For the variance $\sigma^2_{\gamma} $ of the individual specific vectors $\gamma_{i}$ for i=1,...,$N$: 
\begin{align*}
    \pi(\sigma^2_{\gamma} | -) &\sim inv-gamma(a_{\gamma}^*,b_{\gamma}^*), \\
    a_{\gamma}^*&=a_{\gamma} +  \frac{Nk}{2},\\
    b_{\gamma}^*&= b_{\gamma} +  \frac{1}{2}\sum_{i=1}^{N}\gamma_{i}^T\gamma_{i}.
\end{align*}
\item For the error variance parameter
$\sigma^2_{\epsilon}$:
\begin{align*}
    \pi(\sigma^2_{\epsilon}| -) &\sim inv-gamma(a_{\epsilon}^*,b_{\epsilon}^*),\\
    a_{\epsilon}^*&=a_{\epsilon} +  \frac{1}{2} \sum_{i=1}^{N} n_{i},\\
    b_{\epsilon}^*&= b_{\epsilon} +  \frac{1}{2} \sum_{i=1}^{N} (\boldsymbol{Y}_{i} - \boldsymbol{\tilde{m}}_{i})^{T} (\boldsymbol{Y}_{i} - \boldsymbol{\tilde{m}}_{i}),\\
   \boldsymbol{\tilde{m}}_i&=\mathcal{B}_{\beta}(h_i(t,\boldsymbol{\phi_{i}}))^T \boldsymbol{\beta} + \mathcal{B}_{\gamma}(h_{gi}(\boldsymbol{t_{i}},\boldsymbol{\phi_i}))^T \boldsymbol{\gamma}_{i}     \qquad   i=1,...,N.\\
\end{align*}
\end{itemize}
Instead, for the warping parameters, the MH step is still required, and like the priors, it is unaffected by the absence of groups, functioning in the same manner as described in Section \ref{MH step our prior}. 

\section{Further computational details}

\subsection{Convergence Diagnostics for MCMC}
\label{Conv knee}

\begin{figure}[H]
    \centering
    \subfloat[$\beta_2$]{
        \includegraphics[width=4cm]{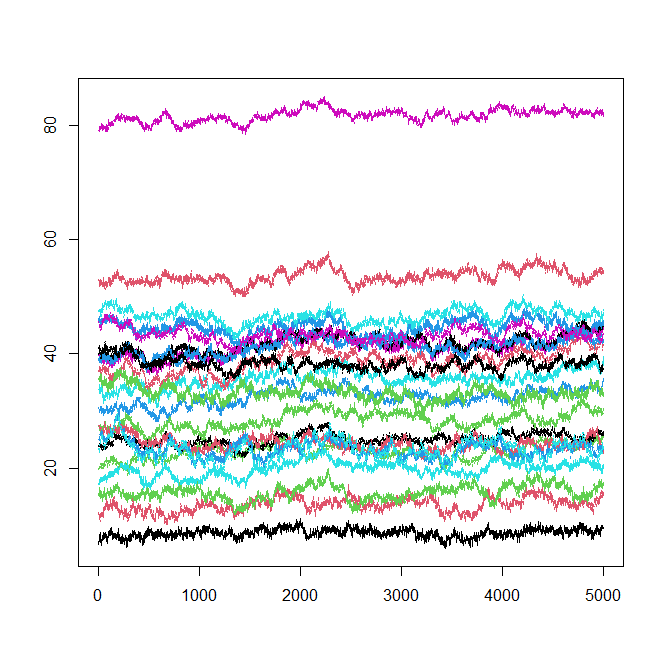}
    }
    \quad
    \subfloat[$\gamma_{1,7}$]{
        \includegraphics[width=4cm]{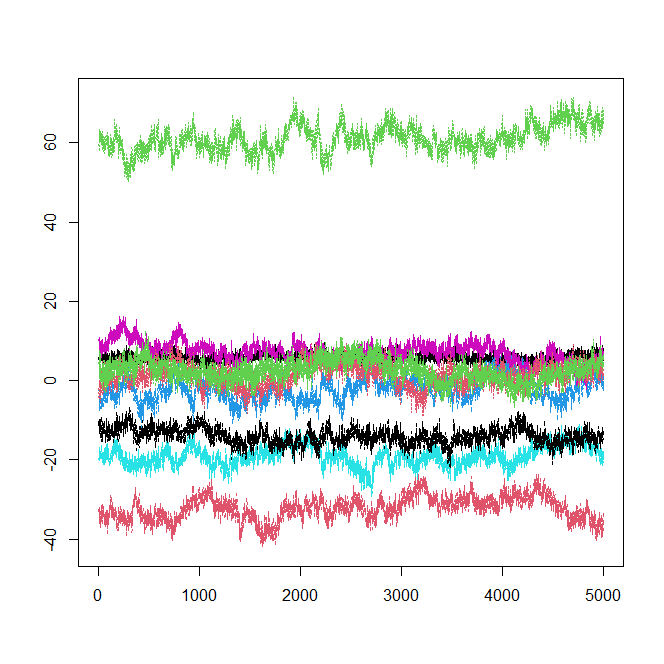}
    }
    \quad
    \subfloat[$\phi_{1,30}$]{
        \includegraphics[width=4cm]{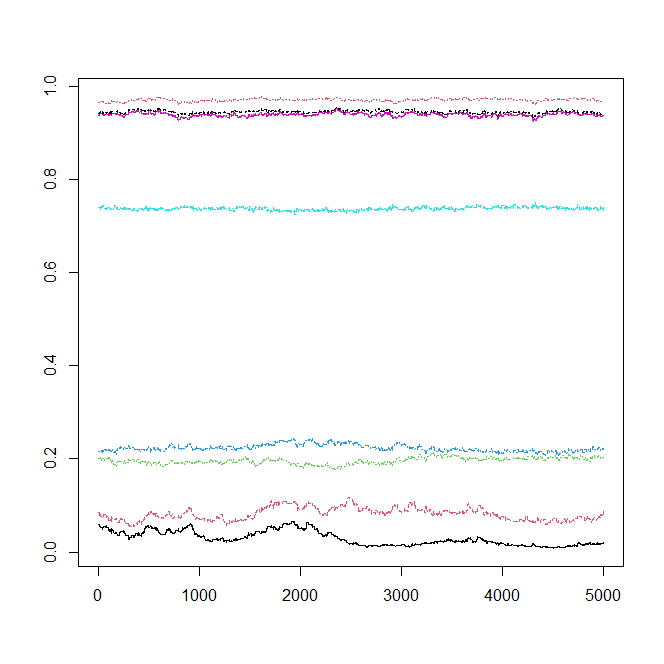}
    }
    \caption{Acceptance rates \& trace plots}
     \label{fig:traceplots}
\end{figure}

In figure \ref{fig:traceplots} are reported the acceptance rates and the trace plots of some paramters of the model. The acceptance rates are within a good range, indicating both good convergence for the warping parameters and thorough exploration of the parameter space. Due to the large number of parameters, only a selection of the trace plots is presented. However, the remaining plots exhibit similar behavior. The presented trace plots do not include the burn-in iterations. As can be seen, the trace plots exhibit a good shape and appear to have reached convergence after a good exploration of the parameter space. 

\begin{figure}[H]
    \centering
     \subfloat[Acceptance rates]{
        \includegraphics[width=4cm]{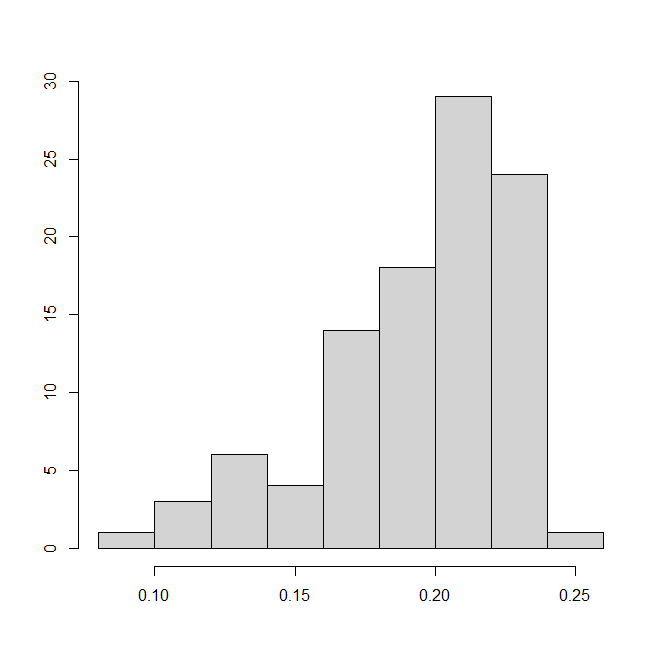}
    }
    \quad
    \subfloat[$\sigma_\gamma^2$]{
        \includegraphics[width=4cm]{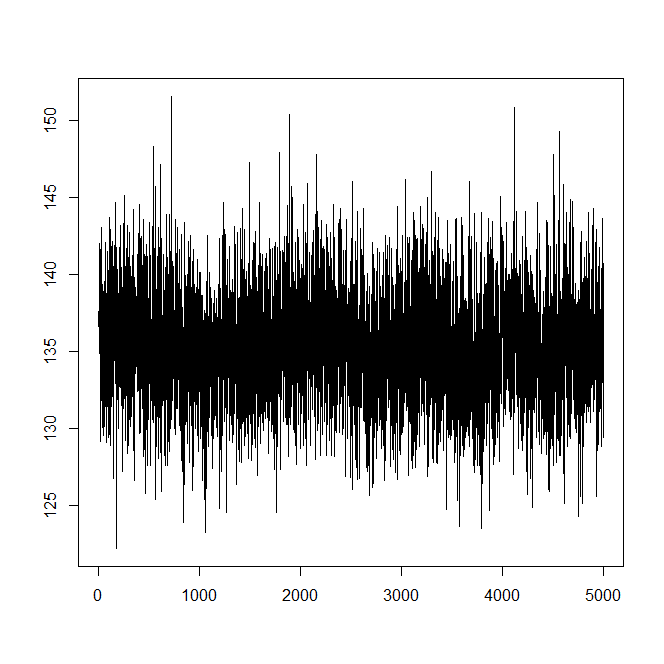}
    }
    \quad
    \subfloat[$\lambda$]{
        \includegraphics[width=4cm]{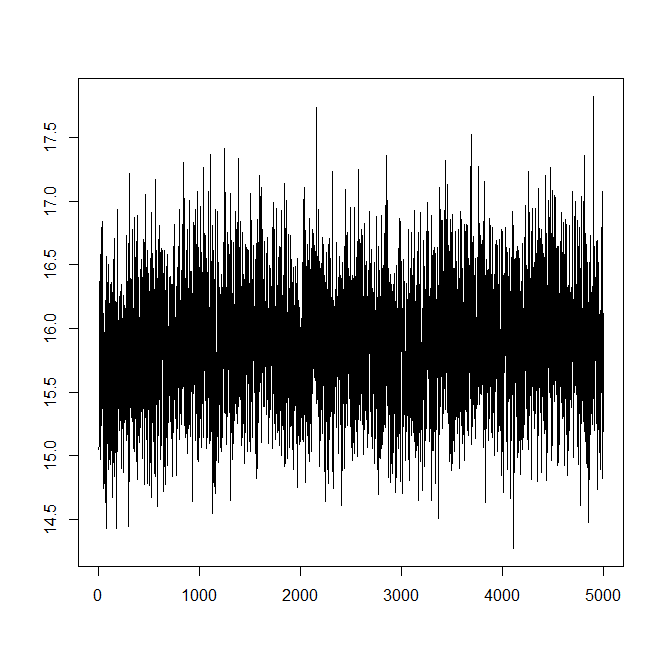}
    }
\end{figure}

\subsection{Examples of aligned curves}
\label{sec:intercept individual term}
\begin{figure}[H]
    \centering
    \subfloat{
        \includegraphics[width=4cm]{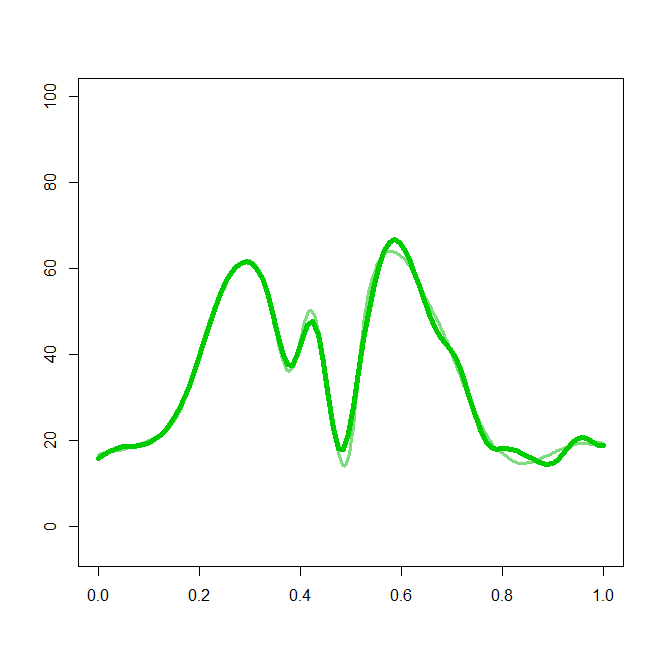}
    }
    \quad
    \subfloat{
        \includegraphics[width=4cm]{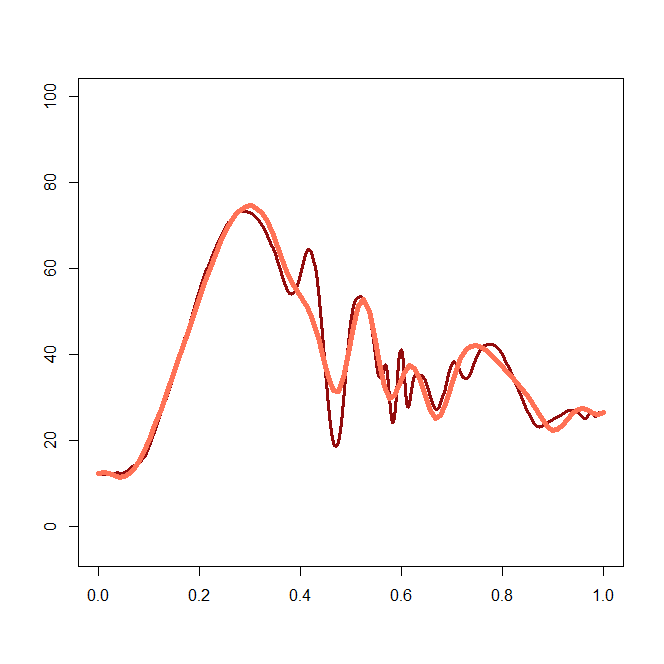}
    }
    \quad
    \subfloat{
        \includegraphics[width=4cm]{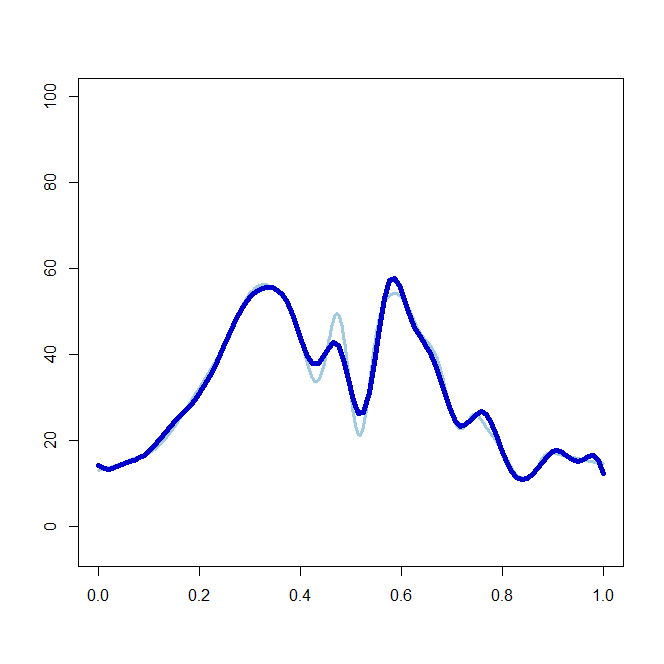}
    }
    \caption{Examples of aligned curves together with the estimated terms \( \mathcal{B}_{\beta}(t)^T \boldsymbol{\beta}_g + \mathcal{B}_{\gamma}(t)^T \boldsymbol{\gamma}_{gi} \) represented by the thickest curves. Left: $y_{1,5}$; Center: $y_{2,30}$; Right: $y_{3,3}$. }
    \label{fig:single_curves}
\end{figure}

In figure \ref{fig:single_curves} we report a few randomly selected aligned curves, obtained using our proposed model, along with their corresponding estimated term \( \mathcal{B}_{\beta}(t)^T \boldsymbol{\beta}_g + \mathcal{B}_{\gamma}(t)^T \boldsymbol{\gamma}_{gi} \), both to show how well it fits the observed data and to illustrate that the behavior near zero is accurately captured by the individual-specific component.

\subsection{Alignment of the data with a too high value of k}
\label{k almost p}
\begin{figure}[H]
    \centering
    \includegraphics[width=4cm]{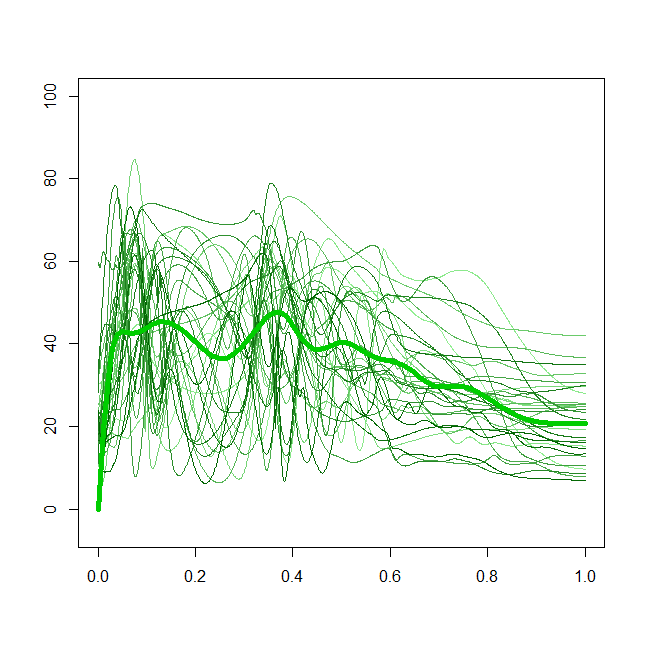}
    \quad
    \includegraphics[width=4cm]{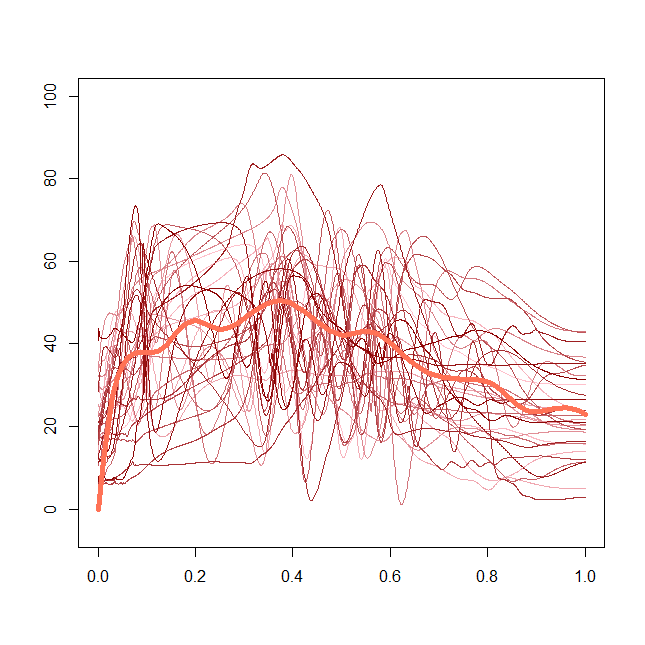}
    \quad
    \includegraphics[width=4cm]{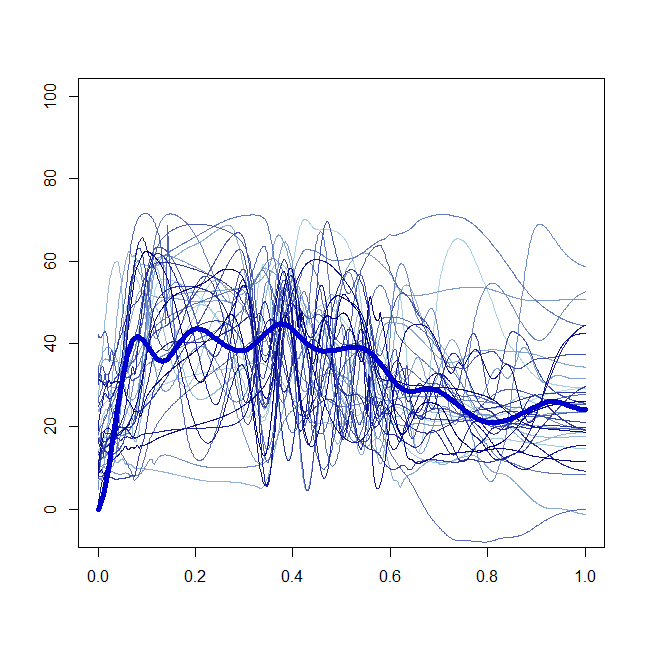}
   \caption{Aligned curves with a too high value of \( k \). Left: Control group; Center: Surgery group; Right: Physiotherapy group.}
    \label{k > p}
\end{figure}

The aligned curves shown in Figure \ref{k > p} were obtained using a model with an excessively high value of \(k\). Specifically, we set \(p = 20\) and \(k = 18\). As can be seen from the plots, the resulting curves are not aligned at all. This, as previously mentioned, occurs due to the smoothed curves fitting the original data perfectly, leaving no room for the alignment step. This also impacts the common term, which is not estimated correctly because the curves are not aligning. For these reasons, a careful selection of vector dimensions is necessary to achieve a satisfactory result.

\subsection{Creation of the second simulated dataset}
\label{Appenddix:Telesca simulated data}
For completeness, in this section we report the model by Telesca and Inoue (2008), which we used as the smoothing component to generate the second simulated dataset. Since we have only used the smoothing part to generate the data, we provide the details of that component only. Instead the warping terms were obtained in the same way as for the first dataset, as previously discussed.
The model we used is: 
\begin{align*}
  y_{i}(t) & = \tilde{m}_{i}(t) + \epsilon_{i}(t)\, \nonumber,\\
  \tilde{m}_{i}(t) &= m_{i}(t) \circ h_{i}(t) = m_{i}(h_{i}(t))\, ,\\
  m_{i}(t) &=  a_i \mathcal{B}_{\beta}(t)^T \boldsymbol{\beta} + c_i\, ,\\
   h_{i}(t) &= \mathcal{B}_h(t)^T \boldsymbol{\phi}_{i}\, ,\\
    \epsilon_{i}(t) &\overset{iid}{\sim}\mathcal{N}(0, \sigma_{\epsilon}^2).
\end{align*}
where $i=1,..,N$ refers to the $i$-th curve, $N$ is the number of curves in the dataset and $\mathcal{B}_{\beta}^T$, $\mathcal{B}_{h}^T$ are $p$ and $q$ dimensional design vectors of cubic B-splines basis functions evaluated at time t.
The prior distributions of the smoothing parameters are:
\begin{itemize}
    \item For the translation parameters $c_i$ of the individual shape function
\begin{align*}
    c_i |c_0,\sigma_c^2  &\overset{iid}{\sim} \mathcal{N} (c_0, \sigma_c^2) \qquad i=1,...,N, \\
    c_0 &\sim \mathcal{N}(m_{c_{0}}, \sigma_{c_0}^2), \\
    \sigma_c^2 &\sim inv-gamma(a_c, b_c).
\end{align*}
\item For the dilation parameters $a_i$ of the individual shape function
\begin{align*}
    a_i |a_0,\sigma_a^2  &\overset{iid}{\sim} \mathcal{N} (a_0, \sigma_a^2) \qquad i=1,...,N, \\
    a_0 &\sim \mathcal{N}(m_{a_{0}}, \sigma_{a_0}^2), \\
    \sigma_a^2 &\sim inv-gamma(a_a, b_a).
\end{align*}
\item For the vector of coefficients $\boldsymbol{\beta}$ of the common shape function
\begin{align*}
    \boldsymbol{\beta} | \lambda &\sim \mathcal{N} (\boldsymbol{0}, \boldsymbol{\Sigma_{\beta}}), \\
    \lambda &\sim inv-gamma(a_{\lambda}, b_{\lambda}).
\end{align*}
where $\boldsymbol{\Sigma_{\beta}}$ is such that $\boldsymbol{\Sigma_{\beta}}^{-1} = \frac{1}{\lambda}\boldsymbol{\Omega}$. $\boldsymbol{\beta}$ can be expressed as a first order random walk:
\begin{equation*}
    \begin{cases}
  \beta_k = \beta_{k-1} + e_k \qquad  e_k \overset{iid}{\sim}\mathcal{N}(0, \lambda),\\
   \beta_1 = 0.
\end{cases}
\end{equation*}
where $k \in \{ 2, ... , p \}$. As a consequence, 
\begin{equation*}
		\boldsymbol{\Omega} = \begin{bmatrix}
			2 &-1 &0 & & &0 \\
			-1 &2 &-1 &0 & &\\
			0 &-1 &\ddots &\ddots &\ddots & \\
			&\ddots &\ddots &\ddots &-1 &0 \\
			& &0 &-1 &2 &-1 \\
			0& & &0 &-1 &1 \\
		\end{bmatrix}	.	
	\end{equation*}
and the random walk variance $\lambda$ can be interpreted as a smoothing parameter for the penalized regression splines.
\item For the variance parameter $\sigma_{\epsilon}^2$ of the error
\begin{equation*}
    \sigma_{\epsilon}^2 \sim inv-gamma(a_{\epsilon}, b_{\epsilon}).
\end{equation*}
\end{itemize}
Specifically, we generate $30$ curves of $300$ observations using the following parameters:\[
c_0 = 3, \quad \sigma_c = 1, \quad \sigma_\phi = 2, \quad a_0 = 1.2, \quad \sigma_a = 0.5, \quad \lambda = 1.3, \quad \sigma_\epsilon = 0.002, \quad p = 14.
\]

\subsection{Telesca and Inoue (2008) alignment of the knee flexion angle dataset}
\label{Appendix: Telesca alignment}
\begin{figure}[H]
    \centering
    \subfloat[Control group]{
        \includegraphics[width=4cm]{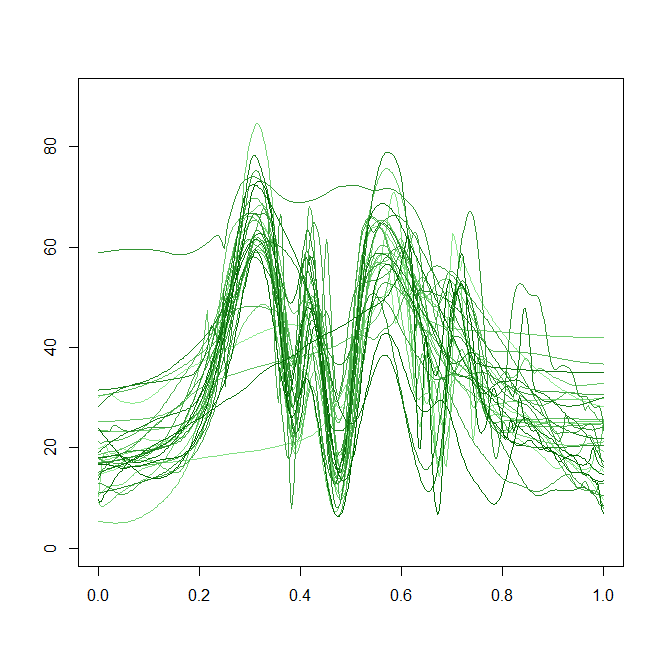}
    }
    \quad
    \subfloat[Surgery group]{
        \includegraphics[width=4cm]{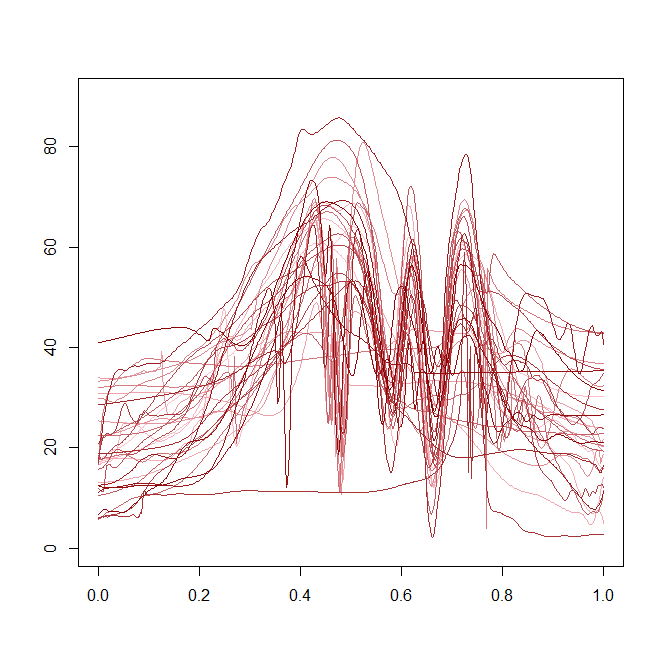}
    }
    \quad
    \subfloat[Physiotherapy group]{
        \includegraphics[width=4cm]{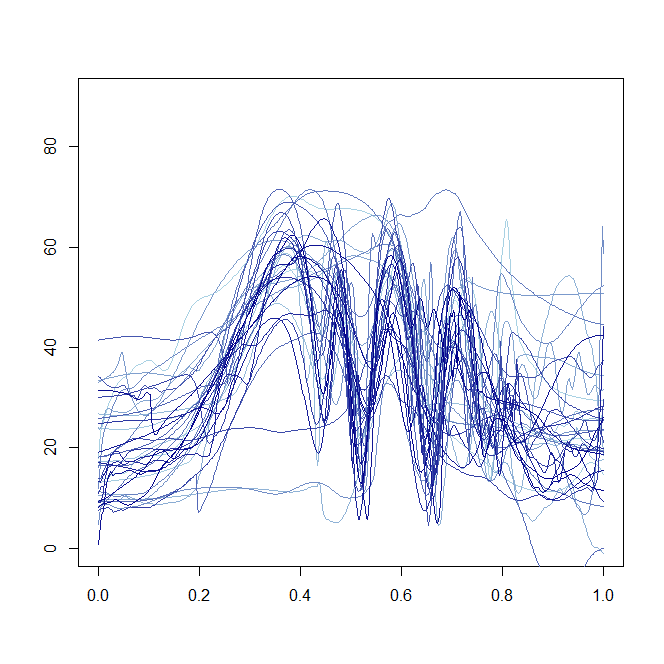}
    }
    \caption{Aligned curves using Telesca and Inoue model}
    \label{fig:Telesca registration}
\end{figure}

\begin{figure}[H]
    \centering
    \subfloat[Control group]{
        \includegraphics[width=4cm]{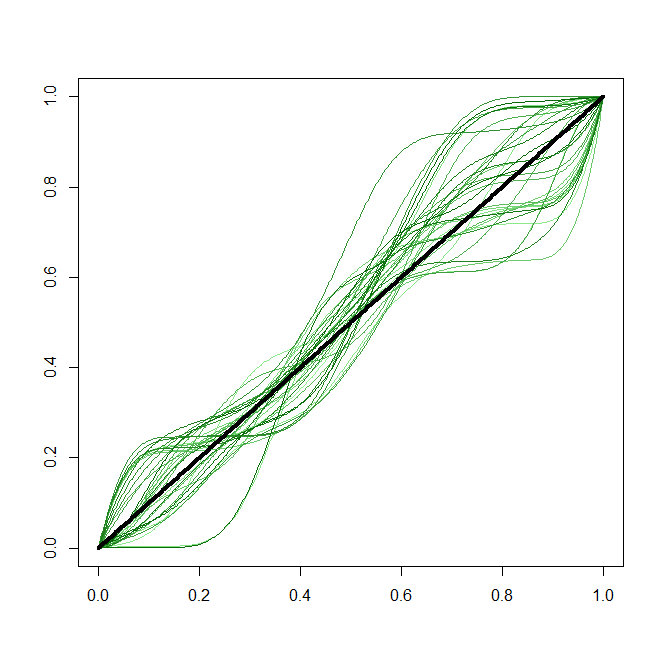}
    }
    \quad
    \subfloat[Surgery group]{
        \includegraphics[width=4cm]{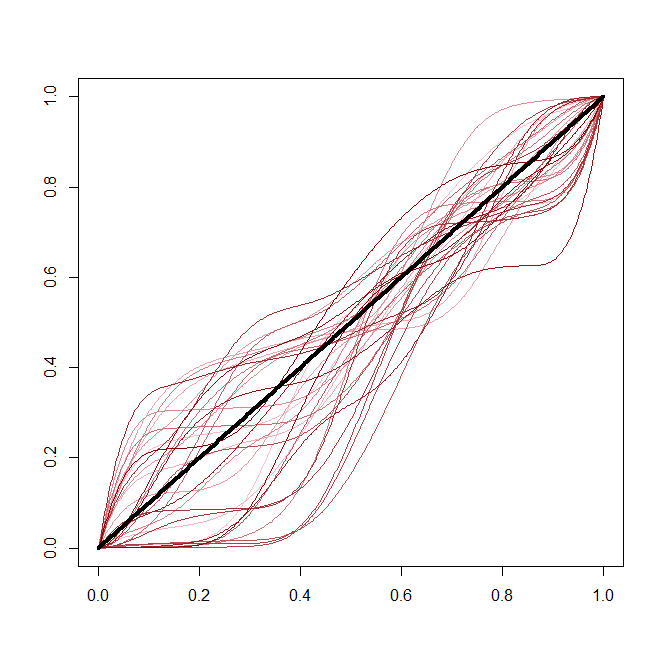}
    }
    \quad
    \subfloat[Physiotherapy group]{
        \includegraphics[width=4cm]{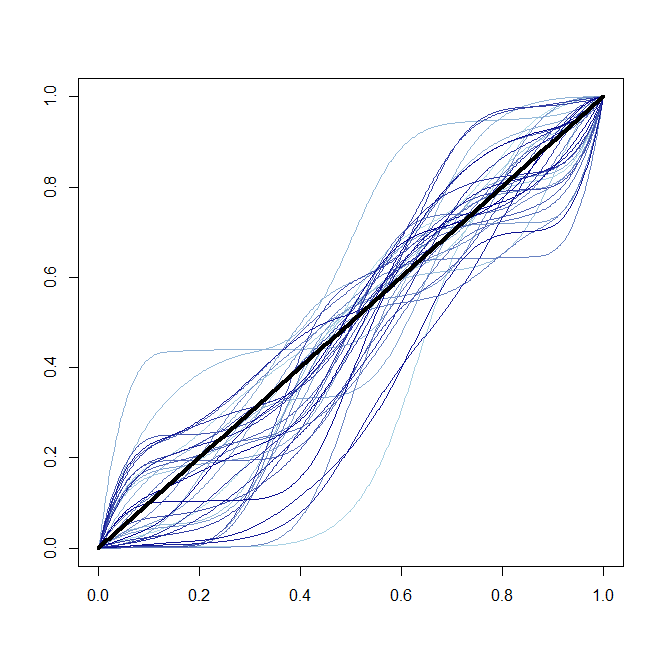}
    }
    \caption{Warping functions using Telesca and Inoue model}
    \label{fig:Telesca warping functions}
\end{figure}

In this section, we present the results obtained by aligning the knee curves using the Telesca and Inoue (2008) model. These results are obtained with 65,000 MCMC iterations, with 50,000 of burn in. It is important to note that, unlike our model where a single run is performed for all three groups simultaneously, in this case three separate runs - one for each group - are required.
The aligned curves are displayed in figure \ref{fig:Telesca registration} and the warping functions in figure \ref{fig:Telesca warping functions}. The curves appear to be aligned, as there is a significant reduction in phase variability. However, there is a noticeable difference compared to our alignment, especially regarding the number of peaks estimated for the three common terms. This is due to the rigidity of the model, which causes many curves to be compressed and deformed in order to estimate the common terms. As a result, curves with fewer peaks are generally aligned correctly, while those with a higher number of peaks or different shapes are significantly altered. In conclusion, although this model is able to align the curves, many individual characteristics are lost or altered during the registration process. This highlights the advantage of using a more flexible alignment method.

\end{document}